\documentclass[fleqn,usenatbib]{mnras}

\usepackage[T1]{fontenc}

\DeclareRobustCommand{\VAN}[3]{#2}
\let\VANthebibliography\thebibliography
\def\thebibliography{\DeclareRobustCommand{\VAN}[3]{##3}\VANthebibliography}

\usepackage{graphicx}	
\usepackage{amsmath,amssymb,bm}	
\usepackage{txfonts}
\usepackage[x11names]{xcolor}
\usepackage{comment}
\usepackage{natbib}
\usepackage{array}
\usepackage{multicol}
\usepackage{subcaption}
\usepackage{arydshln}
\usepackage[normalem]{ulem}

\newcommand{\rexc}{\ensuremath{r_\mathrm{exc}}} 
\newcommand{\RED}[1]{{#1}}
\newcommand{\BUG}[1]{{#1}}
\newcommand{\GREEN}[1]{{#1}}

\defcitealias{Pavan2023}{P23}

\title[Jet injection parameters in BNS mergers]{Role of injection parameters in jet propagation through realistic binary neutron star merger environments}

\author[A. Pavan et al.]{
Andrea Pavan,$^{1,2}$\thanks{E-mail: andrea.pavan@inaf.it (AP)}
Riccardo Ciolfi,$^{1,2}$\thanks{E-mail: riccardo.ciolfi@inaf.it (RC)}
Emma Dreas$^{3,1,4}$
and Jay V. Kalinani$^{5}$
\\
$^{1}$INAF, Osservatorio Astronomico di Padova, Vicolo dell'Osservatorio 5, I-35122 Padova, Italy\\
$^{2}$INFN, Sezione di Padova, Via Francesco Marzolo 8, I-35131 Padova, Italy\\
$^{3}$SISSA, Via Bonomea 265, I-34136 Trieste, Italy\\
$^{4}$INFN, Sezione di Trieste, Via Valerio 2, I-34127 Trieste, Italy\\
$^{5}$Center for Computational Relativity and Gravitation, 
Rochester Institute of Technology, 85 Lomb Memorial Drive, Rochester, New York 14623, USA
}

\date{Accepted XXX. Received YYY; in original form ZZZ}

\pubyear{2025}

\begin{document}
\label{firstpage}
\pagerange{\pageref{firstpage}--\pageref{lastpage}}
\maketitle

\begin{abstract} 
After the first multi-messenger observation of a binary neutron star (BNS) merger powering a short-duration gamma-ray burst (GRB), GW170817-GRB\,170817A, remarkable effort is ongoing to unravel the evolution of the collimated, relativistic outflow (or jet) that was launched during the merger and fed the GRB event, imprinting its angular structure onto the follow-up afterglow signal. Current theoretical models, based on relativistic magneto-hydrodynamic (RMHD) simulations, offer detailed insights into the launch and propagation processes that govern jet evolution. Notably, these simulations point out that jet injection parameters, such as luminosity, magnetization, power decay time scale, and launch time relative to merger, play a crucial role. However, the impact of these parameters is typically investigated within simplified jet propagation environments, lacking a direct connection with a realistic BNS merger aftermath. In this work, we present the first suite of 3D RMHD simulations exploring the influence of such parameters on the propagation of magnetized incipient GRB jets injected into magnetized environments directly imported from the outcome of a general-relativistic MHD BNS merger simulation. Our results demonstrate that, alongside the injection parameters, the BNS merger environment has a central role in shaping the overall jet evolution.
Specifically, under identical jet parameters, the fate of an incipient jet (whether it successfully breaks out or becomes choked) depends strongly on the pro\-perties of \RED{such an} environment.
Further quantitative comparison between realistic and simplified environments reveals major differences, emphasizing the importance of incorporating the former for accurate modeling.
\end{abstract}

\begin{keywords}
MHD –- relativistic processes –- methods: numerical –- stars: jets –- gamma-ray bursts -– neutron star mergers
\end{keywords}


\section{Introduction}

The widely accepted paradigm for gamma-ray bursts (GRBs) attri\-butes these events to electromagnetic (EM) emission arising from collimated, relativistic outflows (`jets') launched by a compact central engine (e.g.,~a hyper-accreting black hole (BH) or a rapidly spinning, highly magnetized neutron star), formed in the aftermath of a massive star's collapse or the merger of a binary neutron star (BNS) or NS-BH system (see, e.g., \citealt{Zhang2018} for a review).
The association with BNS mergers, specifically, was confirmed in 2017 with the first joint detection of gravitational waves (GWs) from the inspiral of two NSs (event named GW170817) and the short-duration GRB\,170817A \citep{LVC-BNS,LVC-MMA,LVC-GRB}.
Follow-up analysis of the accompanying afterglow signal further confirmed that 
the merger launched a jet, emerging with a non-uniform angular structure and powering the GRB. Moreover, the observation angle was constrained between about 15 and 25 degrees relative to the main propagation axis of the jet itself \citep[][]{Lazzati2018,Mooley2018b,Ghirlanda2019,Nakar2021,Mooley2022}.

While the physical connection between BNS mergers and GRBs has been firmly established, the processes driving the jet evolution, from launch by a compact central engine, through propagation in the surrounding post-merger environment, up to formation of an angular structure imprinted on the following EM emission, remain under investigation \citep[][]{Salafia2022}.
Notably, due to the optically thick regime of the system, EM observations cannot directly probe such evolution phases.

Given these observational limitations, a growing effort has been devoted to developing theoretical models that explore the physical conditions at jet launch and during propagation through the post-merger environment, in relation to their final EM signatures.
This effort has involved both analytical/semi-analytical approaches \citep[e.g.,][]{Harrison2018,Lazzati2020,Hamidani2020,Salafia2020,Hamidani2021,Gottlieb2022b,Garcia-Garcia2024,Gutierrez2025} and relativistic (magneto-)hydro\-dynamic simulations \citep[e.g.,][]{Bromberg2018,Duffell2018,Geng2019,Kathirgamaraju2019,Gottlieb2020,Nathanail2020,Gottlieb2021,Murguia2021,Nathanail2021,Nativi2021,Urrutia2021,Gottlieb2022a,Nativi2022,Urrutia2023,Mattia2024,Pais2024}.
In these works, in particular, it has been shown that the properties of the jet at launch (including luminosity, magnetization, and angular profiles), the time delay of the launch itself relative to the merger, and the jet's power decay time scale (linked to the duration of the central engine activity), play a crucial role in shaping the overall jet evolution and its final configuration.

However, while some of the above simulations can capture jet physics with a high level of sophistication, the role of launch conditions on the overall jet evolution is explored using simplified propagation environments, i.e.~surrounding distributions of rest-mass density, pressure, velocity, and magnetic field that are inspired by, but not directly imported from, general-relativistic (GR) BNS merger simulations.

With the aim of providing a more realistic and self-consistent description of the jet propagation process, in  \citet[][henceforth \citetalias{Pavan2023}]{Pavan2023} we developed numerical methods to import magnetized environments directly from the outcome of BNS merger simulations, extending the purely hydrodynamical approach first presented in \cite{Pavan2021} (see also \citealt{Lazzati2021}).
Then, we carried out the first three-dimensional (3D) relativistic magneto-hydrodynamic (RMHD) simulation of an incipient GRB jet manually injected into a `realistic' (i.e.~imported) magnetized BNS merger environment, covering up to 2\,s of jet evolution.
The simulation results, despite being limited to a single fiducial case, demonstrated the crucial importance of employing a realistic magnetized environment and its strong impact on the jet propagation process shaping the final jet angular structure.
In particular, the inhomogeneities and anisotropies characterizing the realistic post-merger ambient medium resulted in a non-axisymmetric final configuration, with implications for the associated EM emission.  
In our previous purely hydrodynamical analysis \citep{Pavan2021}, we also showed that such inhomogeneities and anisotropies, under identical jet injection conditions, result in a significantly different jet breakout time and energy partitioning with respect to the equivalent simulation with a typical simplified environment.

In this paper, we employ the numerical methods developed in \citetalias{Pavan2023} to conduct the first parameter study of GRB jets propagating through realistic BNS merger environments. 
Specifically, we perform the first suite of 3D RMHD simulations of magnetized incipient GRB jets, manually injected with different luminosity, magnetization, power decay time scale, and launch time relative to merger, into a realistic magnetized environment resulting from a 3D GRMHD BNS merger simulation. 
We particularly focus on the impact of such parameters on the jet's dynamics, energetics, and final angular structure, providing quantitative details and comparison with state-of-the-art models in which instead simplified environments are adopted.  
In addition, our simulations extend up to 3\,s after launch, significantly longer than the previous 3D RMHD simulations of GRB jets in BNS merger environments (e.g., \citetalias{Pavan2023} and refs.~above). 

This paper is organized as follows. 
In Section~\ref{sec:setup}, we describe our simulation setup, including numerical methods, data import procedure, computational grid, and jet injection prescription. 
In Section~\ref{sec:5jets}, we present and discuss the results of simulations with a fixed launch time relative to the merger, but different jet injection luminosity, decay time scale, and magnetization.
In Section~\ref{sec:choking_tc}, we examine the results of simulations with different launch time relative to the merger. 
In Section~\ref{sec:angular3s}, we analyze and compare the final jet angular structure across all the simulations in this work.
Finally, in Section~\ref{sec:conclusions}, we summarize our findings and provide an outlook on future directions.

\section{Numerical and physical setup}
\label{sec:setup}

Our simulations are performed using the ideal RMHD module of the \texttt{PLUTO} code \citep[version\,4.4-patch2;][]{Mignone2007-PLUTO1}.
Numerical integration is carried out employing fifth-order piecewise parabolic reconstruction \citep[][]{Mignone2014}, the First ORder CEntred (FORCE) Riemann solver \citep[][]{Toro2006,Mattia2022}, and third-order Runge Kutta time stepping in 3D spherical coordinates $(r,\theta,\phi)$.\footnote{Compared to \citetalias{Pavan2023}, in this work we use a more accurate numerical setup, particularly a less diffusive Riemann solver, in order to increase the effective resolution of our simulations. In Appendix~\ref{app:improvements}, we quantify this increase by comparing the new results with those obtained using a similar setup to \citetalias{Pavan2023}.}
To enforce the free-divergence constraint of the magnetic field, we use Hyperbolic Divergence Cleaning as in \citetalias{Pavan2023}, retaining a cell-centered representation of the primary fluid variables (including the magnetic field) in our simulations.

As initial conditions in \texttt{PLUTO}, we import data directly from the outcome of a BNS merger simulation published in \cite{Ciolfi2020a}, specifically the `B5e15' model (the same model referred to in \citetalias{Pavan2023}). 
The reference merger simulation is performed in 3D GRMHD, setting the initial BNS system with the same chirp mass as estimated for GW170817 \citep[][]{LVC-170817properties}, mass-ratio $\simeq\!0.9$, and magnetic field of the two NS components with a maximum strength of $5\times10^{15}$\,G. 
Specifically, the magnetic field is set as poloidal and internal to each NS.
Furthermore, the adopted equation of state (EOS) is a piecewise-polytropic approximation of the APR4 EOS \citep{Akmal:1998:1804} as implemented in \cite{Endrizzi2016}.
The GR simulation is carried out over a hierarchical 3D-Cartesian grid (maximum resolution $\approx\!250$\,m) extended to $\approx3400$\,km along all axes (equatorial symmetry is imposed on the system orbital plane, i.e., only the $z\ge0$ region is evolved).
During the evolution, an artificial floor with constant density ($6.3\times10^4$\,g/cm$^3$) and pressure is adopted.
We refer the reader to \cite{Ciolfi2020a}, for details on numerical codes and methods.

The BNS merger simulation covers up to $\simeq255$\,ms after merger, leading to the formation of a metastable, magnetized, differentially rotating NS remnant with gravitational mass $M_0\simeq2.596$\,$M_{\sun}$, surrounded by an expanding post-merger environment with anisotropic distribution and non-homologous radial motion.
Within the simulation time span, neither the collapse of the NS remnant into a BH nor the formation of a relativistic jet compatible with GRBs is observed \citep[see][for further physical discussion]{Ciolfi2020a}.
 
To import data from the outcome of the BNS merger simulation into \texttt{PLUTO}, we follow the same procedure as presented in \citetalias{Pavan2023}. In particular:

(i) we remap 3D outputs of rest-mass density, pressure, 3-velocity, and magnetic field onto a uniform Cartesian grid with the same resolution as the hierarchical grid at the radial distance $\rexc=380$\,km;

(ii) we extend the computed values to the $z<0$ region, exploiting the original equatorial symmetry;

(iii) we tilt the imported system by $90\degr$ to arrange the orbital axis along a direction orthogonal to the spherical polar axis (specifically, along the $y$-axis);

(iv) we interpolate the computed values on the \texttt{PLUTO} spherical grid, replacing the original constant floor with a static, unmagnetized `atmosphere' that decays in density and pressure as $r^{-6.5}$ up to the outer radial boundary $r_\mathrm{max}=2.5\times10^6$\,km.

When interpolating data onto the \texttt{PLUTO} spherical grid, we excise the central region at $r<\rexc$, out of which GR effects have no impact on the system dynamics. 
Moreover, we set up the spherical grid with logarithmic and uniform spacing along the radial and angular directions, respectively, and resolution of $768\times256\times512$ cells along $r\in[\rexc,r_\mathrm{max}]$, $\theta\in[0.1,\pi-0.1]$, and $\phi\in[0,2\pi]$, respectively. 
In particular, this resolution allows us to retain the same grid spacing at $\rexc$ as the uniform Cartesian grid in (i).
In Appendix~\ref{app:res_stud}, we also consider a higher grid resolution.

For the time evolution in \texttt{PLUTO}, we account for the (Newtonian) gravitational pull by the central remnant, setting the mass of the latter to $M_0$ as in the reference BNS merger simulation.
Moreover, we relate thermodynamic quantities by means of a Taub-Matthews EOS \citep[][and refs. therein]{Mignone2007}, corresponding to an ideal gas law with adiabatic index $\Gamma_\mathrm{ad}=4/3$ in the ultra-relativistic limit, and $\Gamma_\mathrm{ad}=5/3$ in the non-relativistic one, with a smooth and continuous behavior at intermediate regimes. 
Such an EOS, in particular, allows us to properly account for the non-relativistic environment imported from \cite{Ciolfi2020a}, and for a relativistic jet manually injected into it (see injection recipe in the next Subsection).
We refer the reader to \citetalias{Pavan2023}, for discussion on the impact of our EOS choice, with respect to the simpler $\Gamma_\mathrm{ad}=4/3$ (more appropriate for the treatment of radiation-dominated systems) and to the original EOS employed in the reference BNS merger simulation.

During the time evolution, we assume that the metastable NS remnant eventually collapses into a BH-accretion disk system, which shortly afterward is able to launch an incipient GRB jet along the orbital axis of the progenitor binary.
In particular, we treat the time ($t_\mathrm{c}$) at which such \RED{a} collapse occurs as a free parameter of our simulations.
At $t=t_\mathrm{c}$, then, we introduce a `collapse phase' lasting 30\,ms, during which we take into account (via the corresponding prescription as detailed in \citetalias{Pavan2023}) the decrease in radial pressure gradients in the innermost regions of the domain due to the formation of the central BH-disk.
In particular, we model the temporal trend of such \RED{a} decrease with an exponential decay function, with time scale
\begin{equation}\label{eq:tau}
    \tau = \tau_\mathrm{j} + (\tau_\mathrm{d}-\tau_\mathrm{j})\sin^2{\Theta}\,,
\end{equation}
where $\tau_\mathrm{j}=30$\,ms is the delay time between NS collapse and incipient jet launch, $\tau_\mathrm{d}$ is the accretion time scale of the BH-disk, and $\Theta$ is the angle with respect to the orbital axis of the system (not to be confused with the angle $\theta$ with respect to the spherical coordinates polar axis, which is orthogonal to the orbital axis as a consequence of step (iii) described above).

\subsection{Jet prescription}
\label{sec:jet}

In this work, we account for a variety of incipient GRB jets manually injected via the prescription presented in \citetalias{Pavan2023}, varying the physical parameters of that prescription, such as jet launch time, initial jet luminosity and magnetization (see later), and power decay time \GREEN{scale.}

At the time of launch (i.e. $t_\mathrm{c}+\tau_\mathrm{j}$), at $r=\rexc$, all our jets are set with half-opening angle $\Theta_\mathrm{j}=10\degr$, and initial and terminal Lorentz factors $\Gamma_\mathrm{j}=3$ and $\Gamma_\mathrm{\infty}=300$, respectively. 
For $\Theta\le\Theta_\mathrm{j}$, then, we set uniform radial velocity, namely, 
\begin{equation}
    v^r = c\sqrt{1-\dfrac{1}{\Gamma^2_\mathrm{j}}}\,,
\end{equation}
where $c$ is the light speed, and uniform rotation,
\begin{equation}\label{vphi}
    v^\Phi=\overline{\Omega}\,\rexc\sin{\Theta}\,,
\end{equation}
where $\overline{\Omega}\simeq10$\,rad/s is the jet's angular velocity as calculated in \citetalias{Pavan2023}.
In Eq.~\ref{vphi}, $\Phi$ denotes the azimuthal angle around the jet injection axis (i.e. the $y$-axis, see above).
Since the latter is orthogonal to the spherical polar axis, $\Phi$ is not the azimuthal angle of the spherical coordinates $(r,\theta,\phi)$.
Consequently, $v^\Phi$ turns out to be a combination of the linear velocities $v^\theta$ and $v^\phi$, the latter being computed as (see \citetalias{Pavan2023})
\begin{align}
    v^\theta &= F^{\theta}(\theta,\phi)v^\Phi\,, \\
    v^\phi &= F^\phi(\theta,\phi)v^\Phi\,,
\end{align}
where 
\begin{align}
    F^\theta(\theta,\phi)&=\csc{\theta}\cos{\Phi}\,,\\
    F^\phi(\theta,\phi)&=\sin{\phi}\sin{\Phi}\,,
\end{align}
being $\Phi$ calculated as
\begin{equation}
    \Phi = -\mathrm{sgn}(\cos{\theta})\arccos{\left(\dfrac{\mathrm{sgn}(\cos{\phi}\sin{\theta})}{\sqrt{1+(\cos{\phi}\tan\theta)^{-2}}}\right)}\,.
\end{equation}
The above transformations allow us to inject uniformly rotating jets along both the $y>0$ and $y<0$ directions (`two-sided' jet injection).
For the magnetic field of such jets, we adopt similar profiles as \cite{Geng2019} \citep[see also][]{Marti2015}, that is,
\begin{align}
    B^\Phi &= \dfrac{2 B^{\Phi}_\mathrm{j,m}(\Theta/\Theta_\mathrm{j,m})}{1+(\Theta/\Theta_\mathrm{j,m})^2}\,,\label{eq:Bgeng1}\\ 
    B^r &= B_\mathrm{ratio}B^{\Phi}_\mathrm{j,m}\,,\label{eq:Bgeng2}
\end{align}
where $\Theta_\mathrm{j,m}=0.4\,\Theta_\mathrm{j}$ is the angle at which $B^\Phi$ is maximum (i.e. $B^{\Phi}_\mathrm{j,m}$), and $B_\mathrm{ratio}=0.5$. 
As for the velocity field, the $\theta$- and $\phi$-components of the magnetic field are also derived through transformations of $B^\Phi$; however, due to the pseudo-vector nature of the magnetic field itself, it is necessary to account for a sign change for jet injection along the negative direction, i.e.\footnote{We note that the sign function in Eqs.~\ref{eq:Br}-\ref{eq:Bphi} was not accounted for in \citetalias{Pavan2023}, leading to unphysical north-south asymmetries in the reproduced jet pattern. In the present work, we have solved this problem, and the differences with respect to \citetalias{Pavan2023} are discussed in Appendix~\ref{app:improvements}.}
\begin{align}
    B^r &= \mathrm{sgn}(y)B^r\,,\label{eq:Br} \\
    B^\theta &= \mathrm{sgn}(y)F^\theta(\theta,\phi)B^\Phi\,,\label{eq:Bth} \\
    B^\phi &= \mathrm{sgn}(y)F^\phi(\theta,\phi)B^\Phi\,.\label{eq:Bphi}
\end{align}
Finally, the rest-mass density and pressure profiles, $\rho(\Theta)$ and  $P(\Theta)$, respectively, are derived \RED{following the calculations in Appendix~B1 of \citetalias{Pavan2023}}, for specified values of initial jet luminosity and relative contribution of magnetic luminosity, $L_\mathrm{j}$ and $\Sigma_\mathrm{j}$, respectively, \BUG{being}\footnote{\BUG{Here, we adopt the correct definition of $L_\mathrm{j}$, rectifying the version reported in \citetalias{Pavan2023} where $T^{00}$ was used instead of $T^{0r}$.}}  
\begin{equation}
    \BUG{L_{\mathrm{j}}/2=\int_0^{\Theta_\mathrm{j}}\int_0^{2\pi}T^{0r}(\rexc,\Theta,\Phi)\,c\,\rexc^2\sin{\Theta\mathrm{d}\Theta\mathrm{d}\Phi},}
\end{equation}
\BUG{with $T^{0r}$ the time-radius component of the energy-momentum tensor and}
\begin{equation}\label{eq:Lmag_tot}
    \Sigma_\mathrm{j} \equiv \dfrac{L_\mathrm{j,B}}{L_\mathrm{j}}\,,
\end{equation}
where $L_\mathrm{j,B}$ is the magnetic luminosity of the jet, i.e. $L_\mathrm{j}$ with purely hydrodynamic contributions set to \GREEN{zero.} \RED{In particular,} $L_\mathrm{j,B}\propto(B^{\Phi}_\mathrm{j,m})^2$. Thus, $L_\mathrm{j,B}$ and $\Sigma_\mathrm{j}$ are not free parameters of our prescription, but de\-rived once $B^{\Phi}_\mathrm{j,m}$ and $L_\mathrm{j}$ are set.

After the jet is launched, the above prescription is applied continuously over time, introducing a time dependence for the jet variables.
Specifically, as in \citetalias{Pavan2023}, we adopt the following time trends:
\begin{align}
    \rho,v^{\Phi} &= \mathrm{const.}\,,\\
    v^r(t) &= \dfrac{\Gamma_\mathrm{j}v^re^{-t/2\tau_\mathrm{d}}}{\Gamma(t)}\,,\\ 
    B^{\Phi}(t) &= B^{\Phi}\sqrt{\dfrac{v^r}{v^r(t)}}e^{-t/2\tau_\mathrm{d}}\,,\\ 
    B^r(t) &= B^r\sqrt{\dfrac{v^r}{v^r(t)}}e^{-t/2\tau_\mathrm{d}}\,,\\ 
    P(t) &= \dfrac{1}{4}\left[\rho c^2(h^*(t)-1)-b^2(t)\right]\,,
\end{align}
where
\begin{align}
    \Gamma(t)=\sqrt{1+\left(\Gamma_{\mathrm{j}}v^re^{-t/2\tau_{\mathrm{d}}}/c\right)^2}\,,  \label{Gm_t}
\end{align}
and
\begin{equation}
    h^*(t)=\dfrac{h^*_\mathrm{0}\Gamma_{\mathrm{j}}e^{-t/2\tau_{\mathrm{d}}}}{\Gamma(t)}\,, \label{h*_t}
\end{equation}
being $h^{*}_\mathrm{0}\equiv\Gamma_{\infty}/\Gamma_\mathrm{j}$, $b^2$ the magnetic field strength squared measured in the proper frame, that is,
\begin{equation}
    b^2 = \dfrac{1}{4\pi}\left[\dfrac{B^2}{\Gamma^2}+\dfrac{(\vec{v}\cdot\vec{B})^2}{c^2}\right]\,,    
\end{equation}
and $\tau_\mathrm{d}$ the same time scale of BH-disk as in Section~\ref{sec:setup}. 
In particular, the above trends result in a jet luminosity that decays over time, with a faster decay occurring for smaller values of \GREEN{$\tau_\mathrm{d}$}.

\section{Impact of jet parameters}
\label{sec:5jets}

In this Section, we discuss the results of simulations performed using the above numerical and physical setup, in which we adopt a launch time of $385$\,ms after merger and vary the parameters $L_\mathrm{j}$, $B^{\Phi}_\mathrm{j,m}$, and $\tau_\mathrm{d}$. The impact of a different launch time will be addressed in Section~\ref{sec:choking_tc}.

In all our simulations, we import data for the realistic BNS merger environment at $t_0=155$\,ms after merger, that is just before the environment material expands beyond the outer boundary of the original computational grid (thus avoiding any loss of physical information; see \citetalias{Pavan2023}).
Therefore, from $t_0$ to 355\,ms after merger (i.e. $t_\mathrm{c}$), we need to evolve such \RED{an} environment in \texttt{PLUTO}, setting appropriate boundary condition on the $r=\rexc$ excision surface.
Following \citetalias{Pavan2023}, we set the initial angular distribution of each primary fluid variable, multiplied by a time function best fitting the temporal trend of the respective angle-averaged quantity (see \citetalias{Pavan2023} for details).
The resulting physical evolution is then refined via the substitution method developed in \citetalias{Pavan2023}, which allows us to exploit the results of the reference BNS merger simulation up to the last available time of the latter, i.e. 255\,ms after merger.
After reaching the envisage $t_\mathrm{c}$ and carrying out the next 30\,ms collapse phase (see Section~\ref{sec:setup}), we start jet injection.
We refer the reader to Figure~7 in \citetalias{Pavan2023}, for the physical properties of the BNS merger environment at $385$\,ms after merger.

In the present paper, jet propagation is reproduced until 3\,s after launch, covering the following physical steps.

(1) \textit{Propagation within \RED{a} dense environment}

At early times, the jet drills through the high-density regions of the surrounding BNS merger environment. 
This occurs through forward and reverse shocks at the jet front, heating the inflowing material at the expense of the jet's own kinetic energy.
The heated material is then diverted sideways (as its pressure exceeds that of the surrounding unshocked medium), forming a pressured cocoon around the jet, which keeps it collimated during propagation. 
Meanwhile, thermomagnetic energy is converted into kinetic form within the jet, allowing the material at the front to accelerate and apply the necessary ram pressure to sustain the propagation itself \citep[see][for a more in-depth physical discussion]{Bromberg2011}.
During the above process, anisotropies in the realistic environment cause the jet to snake along regions of lower density, developing turbulence and asymmetries at the front (where the jet-environment interplay is most intense; \citetalias{Pavan2023}).
The presence of magnetic fields, on the other hand, reduces mixing between the jet and surrounding environment, allowing the former to retain higher stability and collimation during propagation \citep[][]{Mignone2010,Gottlieb2020,Gottlieb2021}.\footnote{We remark that Poynting flux-dominated jets are prone to current-driven instabilities, e.g. kink, which can make the jet configuration globally \GREEN{unstable} \citep[][]{Bromberg2016}.}.

(2) \textit{Jet breakout}

If the above drilling is successful, the jet breaks out of the BNS merger environment and begins to propagate into the rarefied outer medium.
Due to the strong density and pressure gradients upstream, the jet front undergoes rapid acceleration, reaching ultra-relativistic velocities (the bulk Lorentz factor can increase by a factor $\simeq10$; see Figure~8 in \citetalias{Pavan2023}).
Simultaneously, it undergoes lateral expansion, widening the jet's opening angle \citep[][]{Mizuta2013}. 
As a result, any asymmetry previously generated by interaction with the realistic environment is spread out over larger angular scales.
On the other hand, toward the jet base, the injected material \RED{leads to the formation of} a highly collimated, relativistic tail.

(3) \textit{Evolution to full detachment}

When the injection stops or drops enough to become negligible, the tail of the jet also begins to detach from the BNS merger environment, accelerating into the rarefied medium (in \citetalias{Pavan2023}, this occurs at $\simeq\!1$\,s after launch, when $L_\mathrm{j}(t)$ has decreased by a factor $\simeq\!30$; see third rendering from the left in Figure~8 of that work).
Notably, the tail material propagates into the previously drilled channel, converting much of the injected energy into kinetic form, up to speeds even higher than those at the jet front.
Consequently, the jet material at the bottom approaches the top, forming a kinetically dominated bullet (hereafter referred to as the `jet head').

(4) \textit{Free expansion towards ballistic phase}

Once the jet head has completely detached, it expands freely into the rarefied medium, converting the remaining thermomagnetic energy into kinetic form. 
As a result, the bulk Lorentz factor increases towards the local saturation value (the latter depending on $\Gamma_\infty$ at launch and the amount of energy expended in drilling).
At the same time, energy redistribution due to internal interactions can reduce the local maxima of the Lorentz factor, transferring energy to slower fluid elements.
Long enough after the full detachment (and beyond the 3\,s after launch covered in the present work), the expansion itself will eventually become `ballistic', that is, all the energy is in kinetic form and the angular structure remains frozen (see further discussion in \citealt{Dreas2025}).
\begin{table*}
    \centering
    \caption{Physical models investigated in this paper. For each model, labeled as in the first column on the left, we list the fixed jet parameters, i.e. half-opening angle ($\Theta_\mathrm{j}$), and initial and terminal Lorentz factors ($\Gamma_\mathrm{j}$ and $\Gamma_{\infty}$, respectively), and the variable jet launch time (sum of the NS collapse time, $t_\mathrm{c}$, and $\tau_\mathrm{j}=30$\,ms; see Section~\ref{sec:setup}), initial jet luminosity ($L_\mathrm{j}$) and maximum toroidal magnetic field ($B^{\Phi}_\mathrm{j,m}$), and decay time-scale ($\tau_\mathrm{d}$). In addition, in the last two columns on the right, we list the initial jet magnetic luminosity ($L_\mathrm{j,B}$) and relative contribution to $L_\mathrm{j}$ (i.e. $\Sigma_\mathrm{j}$; Eq.~\ref{eq:Lmag_tot}).
    We recall that other fixed jet parameters of our prescription include the initial poloidal-to-toroidal magnetic field ratio ($B_\mathrm{ratio}=0.5$) and angular distance from the jet injection axis at which $B^{\Phi}=B^{\Phi}_\mathrm{j,m}$ \GREEN{(i.e. $\Theta_\mathrm{j,m}=0.4\,\Theta_\mathrm{j}$)}.
    }
    \label{tab:1}
    \begin{tabular*}{2.084\columnwidth}{@{\extracolsep{\fill}}lccccccccc@{}}
        \hline
        \hline
        \noalign{\vskip 1mm}
         & \multicolumn{3}{c}{Fixed Parameters} & \multicolumn{4}{c}{Variable Parameters} & \\
        \noalign{\vskip 1mm}
        \cline{2-4}
        \cline{5-8}
        \noalign{\vskip 1mm}
        Model & $\Theta_\mathrm{j}\,[\degr]$ & $\Gamma_\mathrm{j}$ & $\Gamma_{\infty}$ & $(t_\mathrm{c}+\tau_\mathrm{j})$\,$[\mathrm{ms}]$ & $L_\mathrm{j}$\,$[\mathrm{erg/s}]$ & $B^{\Phi}_\mathrm{j,m}$\,$[\mathrm{G}]$ & $\tau_\mathrm{d}$\,$[\mathrm{ms}]$ & $L_\mathrm{j,B}$\,$[\mathrm{erg/s}]$ & $\Sigma_\mathrm{j}$\\
 	\noalign{\vskip 0.6mm}
 	\hline
 	\noalign{\vskip 1mm} 
 	(a) & 10 & 3 & 300 & 385  & \BUG{$8.95\times10^{51}$} & $1.56\times10^{13}$ & 300 & \BUG{$1.07\times10^{50}$} & \BUG{$1.20\times10^{-2}$}\\
 	(b) & 10 & 3 & 300 & 385  & \BUG{$5.37\times10^{51}$} & $1.21\times10^{13}$ & 500 & \BUG{$0.64\times10^{50}$} & \BUG{$1.20\times10^{-2}$}\\
 	(F) & 10 & 3 & 300 & 385  & \BUG{$5.37\times10^{51}$} & $1.21\times10^{13}$ & 300 & \BUG{$0.64\times10^{50}$} & \BUG{$1.20\times10^{-2}$}\\
 	(c) & 10 & 3 & 300 & 385  & \BUG{$5.37\times10^{51}$} & \BUG{$1.38\times10^{13}$} & 300 & \BUG{$0.84\times10^{50}$} & \BUG{$1.56\times10^{-2}$}\\
 	(d) & 10 & 3 & 300 & 385  & \BUG{$4.12\times10^{51}$} & $1.21\times10^{13}$ & 300 & \BUG{$0.64\times10^{50}$} & \BUG{$1.56\times10^{-2}$}\\
 	(e$_1$) & 10 & 3 & 300 & 385  & \BUG{$3.69\times10^{51}$} & $0.62\times10^{13}$ & 300 & \BUG{$0.17\times10^{50}$} & \BUG{$0.46\times10^{-2}$}\\
 	(e$_2$) & 10 & 3 & 300 & 185  & \BUG{$3.69\times10^{51}$} & $0.62\times10^{13}$ & 300 & \BUG{$0.17\times10^{50}$} & \BUG{$0.46\times10^{-2}$}\\  
        \noalign{\vskip 0.5mm}  
 	\hline
        \hline
        \noalign{\vskip 0.5mm}
    \end{tabular*}
    {\raggedright \textit{Note}: models (e$_1$) and (e$_2$) are analyzed in Section~\ref{sec:choking_tc}. \par}
\end{table*}

In \citetalias{Pavan2023}, we analyzed the above process mostly in a qualitative way.
Moreover, we focused our analysis on a single jet-environment system.
In the following, we present a more quantitative and systematic investigation based on the evolution and comparison of a variety of such systems.

\subsection{Fiducial model}
\label{sec:fiducial}

We first discuss the results of our fiducial simulation performed setting $L_\mathrm{j}$, $B^{\Phi}_\mathrm{j,m}$, and $\tau_\mathrm{d}$ to \BUG{$5.37\times10^{51}$\,erg/s}, $1.21\times10^{13}$\,G, and 300\,ms, respectively.
This configuration, in particular, corresponds to a magnetic luminosity \BUG{of $L_\mathrm{j,B}=0.64\times10^{50}$\,erg/s}, thus \BUG{$\Sigma_\mathrm{j}=1.20\times10^{-2}$}.
Along with $t_\mathrm{c}=355$\,ms after merger, such \RED{a} configuration defines our fiducial model, hereafter labeled (F), which is used in this paper as a reference to set up the other models and make quantitative comparisons.
In Table~\ref{tab:1}, we list the configuration adopted for each model, to be presented throughout the paper.

\begin{figure}
    \centering 
    \includegraphics[width=\columnwidth]{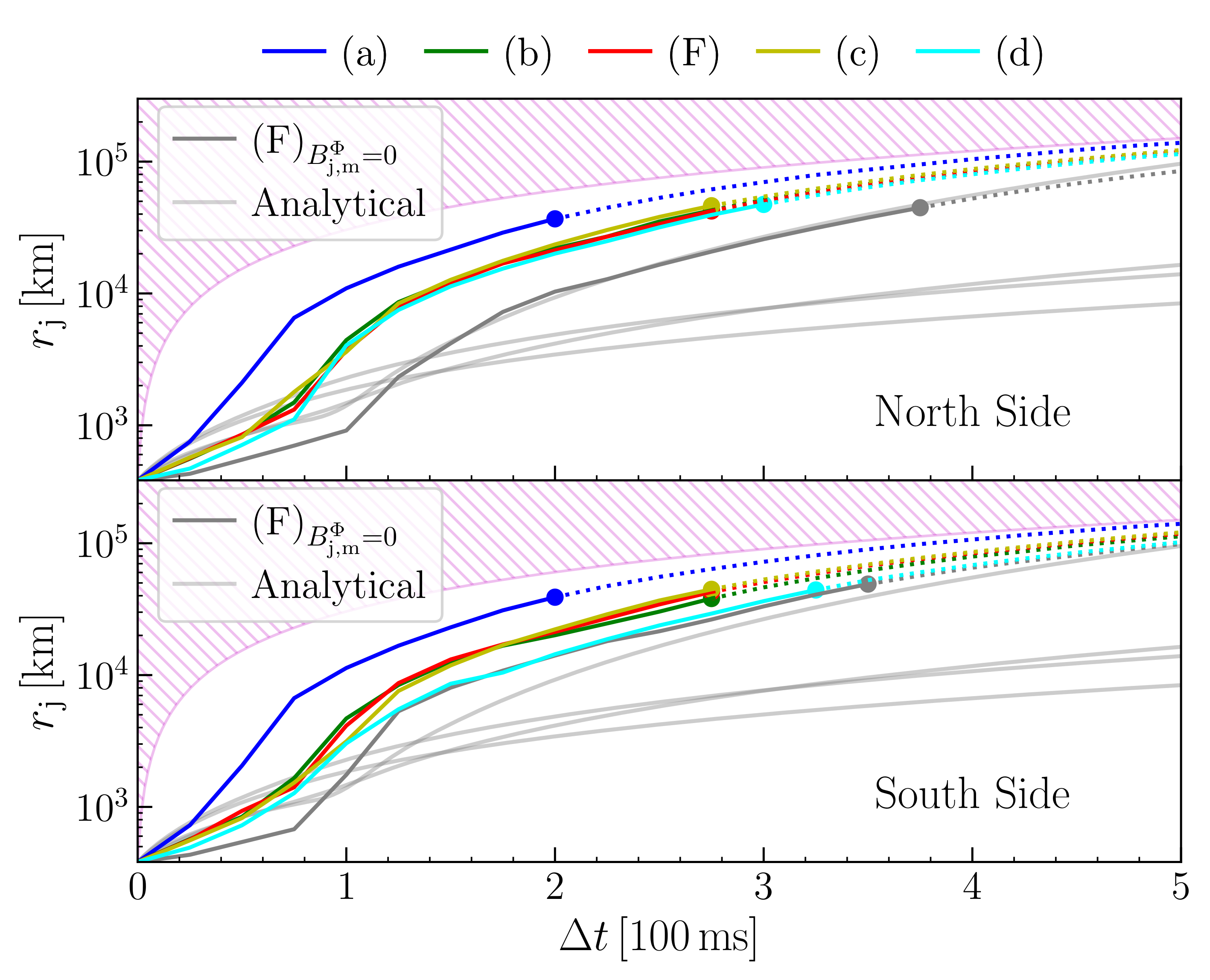}
    \caption{Radial distance ($r_\mathrm{j}$, log scale) traveled by the jet front as a function of time, from launch to 500\,ms later, in the north and south sides ($\phi\in[0,\pi]$ and $\phi\in[\pi,2\pi]$, respectively) of the \texttt{PLUTO} domain, for different models. 
    The solid/dotted colored lines indicate trends before/after the jet breakout (dot markers). 
    The hatched region indicates where causality is violated, i.e. $r_\mathrm{j}>\rexc+c\,\Delta t$. See text for additional details and discussion.
    }
    \label{fig:fig1}
\end{figure}

Figure~\ref{fig:fig1} shows the radial distance ($r_\mathrm{j}$, in log scale) traveled by the jet front as a function of time, from launch to 500\,ms later ($\Delta t$ being the elapsed time since launch), in the north and south sides (namely $\phi\in[0,\pi]$ and $\phi\in[\pi,2\pi]$, respectively) of the \texttt{PLUTO} spherical domain, for different models.
Specifically, $r_\mathrm{j}$ is the radial distance at which, at a given computational time (outputs saved every 25\,ms), going from the outer radial boundary $r_\mathrm{max}$ (specified above) toward $\rexc=380$\,km along each radial direction, we first encounter shocked material interacting with the jet with velocity larger than \GREEN{$0.8\,c$}.
The time trends of such \RED{a} distance are shown as solid/dotted colored lines before/after the jet breakout (dot markers), which occurs when $r_\mathrm{j}$ first exceeds the radial extension of the (expanding) BNS merger environment.
Additional light grey profiles refer instead to analytical estimates discussed below. 
Finally, the hatched region in the Figure indicates where causality is violated, i.e. $r_\mathrm{j}>\rexc+c\,\Delta t$.

Focusing on model (F), we note that $r_\mathrm{j}$ grows immediately after the jet launch, both in the north and south sides, indicating prompt propagation of the jet into the BNS merger environment.
According to analytical modeling by \cite{Gottlieb2022b}, this is typical of a `strong jet' regime, in which (see Subsections~3.1-3.2, 4.4 and Tables~1, 4 of the cited work)
\begin{equation}\label{eq:eta_crit}
    \eta_\mathrm{crit} \equiv N_\mathrm{crit}\left[\dfrac{L_\mathrm{j}\Theta_\mathrm{j}^{-2}\left(t_\mathrm{c}+\tau_\mathrm{j}\right)}{\rho_\mathrm{e}\rexc^3v_\mathrm{e}^2}\right]^{1/2}>1\,,  
\end{equation}
where \GREEN{$N_\mathrm{crit}=0.5$,} $\rho_\mathrm{e}$ and $v_\mathrm{e}$ are, respectively, the rest-mass density and velocity of the BNS merger environment at $r=\rexc$ and $\Theta=0$ at \RED{the time of jet launch ($t_\mathrm{c}+\tau_\mathrm{j}$).}
In model~(F), specifically, we estimate \BUG{$\eta_\mathrm{crit}\!\simeq\!35$} \GREEN{both in the north and south sides, consistent with the above regime.}

To directly compare our simulation results with analytical predictions by \cite{Gottlieb2022b}, we first note that in the current model (F), at the time the jet is launched, the local magnetization, i.e.~the ratio of Poynting flux to total hydrodynamic energy flux, 
\begin{equation}\label{eq:sigma_small}
    \sigma = \dfrac{b^2}{4\pi\rho hc^2}\,,
\end{equation}
where \RED{$h=1+\epsilon+P/\rho$} is the specific enthalpy ($\epsilon$ being the specific internal energy), has a maximum value within the jet of $\sigma_{\max}\simeq1.52$.
However, the above analytical predictions can be applied only to jets that are purely hydrodynamic or at most weakly magnetized, i.e. $\sigma_{\max}\lesssim0.1$ \citep[][and refs. therein]{Gottlieb2022b}. 
Therefore, for the purpose of direct comparison, we consider here a simulation similar to model (F) but null $B^{\Phi}_\mathrm{j,m}$, i.e. unmagnetized jet at injection. 
\RED{In this simulation, moreover, the initial jet luminosity is lowered to $L_\mathrm{j}=3\times10^{51}$\,erg/s, allowing us to enforce a sub-relativistic regime at the jet front throughout the system evolution (see below).} Note that in this model, due to the magnetic field of the BNS merger environment, the relativistic outflow gains $\sigma_{\max}\simeq0.28$, compatible with a weak magnetization regime.
The time trends of $r_\mathrm{j}$ in this model are shown in darker gray in Figure~\ref{fig:fig1}.
Analytical trends based on \citet{Gottlieb2022b}, shown instead in lighter gray, have been calculated assuming an analytical (unmagnetized) jet propagation environment, with homologous radial expansion, and power-law rest-mass density distribution, isotropic at angles $\Theta\le\Theta_\mathrm{j}$ (anisotropies outside the initial jet opening angle are indeed of little relevance to jet propagation; see Subsection~4.4 in the cited work).
Specifically, we have assumed the following profiles:
\begin{align}
    \rho(r) &= \rho_\mathrm{e}\left(\dfrac{r}{\rexc}\right)^{-\alpha}\left(\dfrac{t}{t_\mathrm{c}+\tau_\mathrm{j}}\right)^{-\alpha}\quad;\quad\Theta\le\Theta_\mathrm{j}\,,\\
    v(r) &= v^r = v_\mathrm{e}\left(\dfrac{r}{\rexc}\right)\quad;\quad v_\mathrm{e}\le v\le v_\mathrm{e,max}\,,
\end{align}
where $\alpha\!\in\!\mathbb{N}$ is constant, $\rho_\mathrm{e}\!\simeq\!10^8$\,g/cm$^3$\GREEN{, $v_\mathrm{e}\!\simeq\!0.05\,c$}, and $v_\mathrm{e,max}\!\simeq\!0.48\,c$ is the maximum velocity of the BNS merger environment at the time of jet launch in the comparison numerical simulation.
In particular, \RED{we have calculated analytical trends} for $\alpha=1,2,3,4$.
In addition, we have assumed that a uniform jet is injected at $r=\rexc$ into the analytical environment, at the same time after merger as in the comparison simulation, \RED{with a constant luminosity equal to the initial value adopted in the simulation itself (i.e.~$L_\mathrm{j}$).}
Finally, we have made use of Eqs.\,25, 44-45 in \cite{Gottlieb2022b} to calculate, for a given $\alpha$, the analytical time trend of $r_\mathrm{j}$ and \RED{the} corresponding jet breakout time, within a sub-relativistic \GREEN{regime} consistent with the results of our \GREEN{simulation}.

Comparing the above analytical trends with the null $B^{\Phi}_\mathrm{j,m}$ simulation, we find very good agreement for $\alpha\!=\!4$ (especially in the north side).
This is consistent with the power-law index typically employed for the description of BNS merger environments \citep[][and refs. therein]{Gottlieb2022b}.
In terms of jet breakout, however, the $\alpha=4$ analytical case predicts $\simeq514$\,ms after launch (estimation performed using Eq.~45 in the cited paper, where the parameter $\Tilde{E}_\mathrm{d}$, defined therein, is $\simeq15.5$ in the present comparison), while our simulation gives $\simeq375$\,ms and $\simeq350$\,ms after launch in the north and south sides, respectively.\footnote{We stress that these differences are found under the assumption of a constant jet injection luminosity in the analytical case.
A jet luminosity that decays over time (as in our simulation) would increase the propagation time of jet, resulting in even larger differences.}
The faster breakout of the latter case, despite the nearly identical time trend of $r_\mathrm{j}$ (i.e.~jet advancement speed), results from the lower expansion rate of the realistic propagation environment compared to a simple homologous model.

The above comparison demonstrates that accurate predictions of jet breakout times in BNS merger context require a realistic propagation environment, particularly rest-mass density and velocity profiles obtained self-consistently in a GR simulation of the BNS merger itself (confirming previous results by \citealt{Pavan2021}).

\begin{figure*}
    \centering \includegraphics[width=1.7\columnwidth]{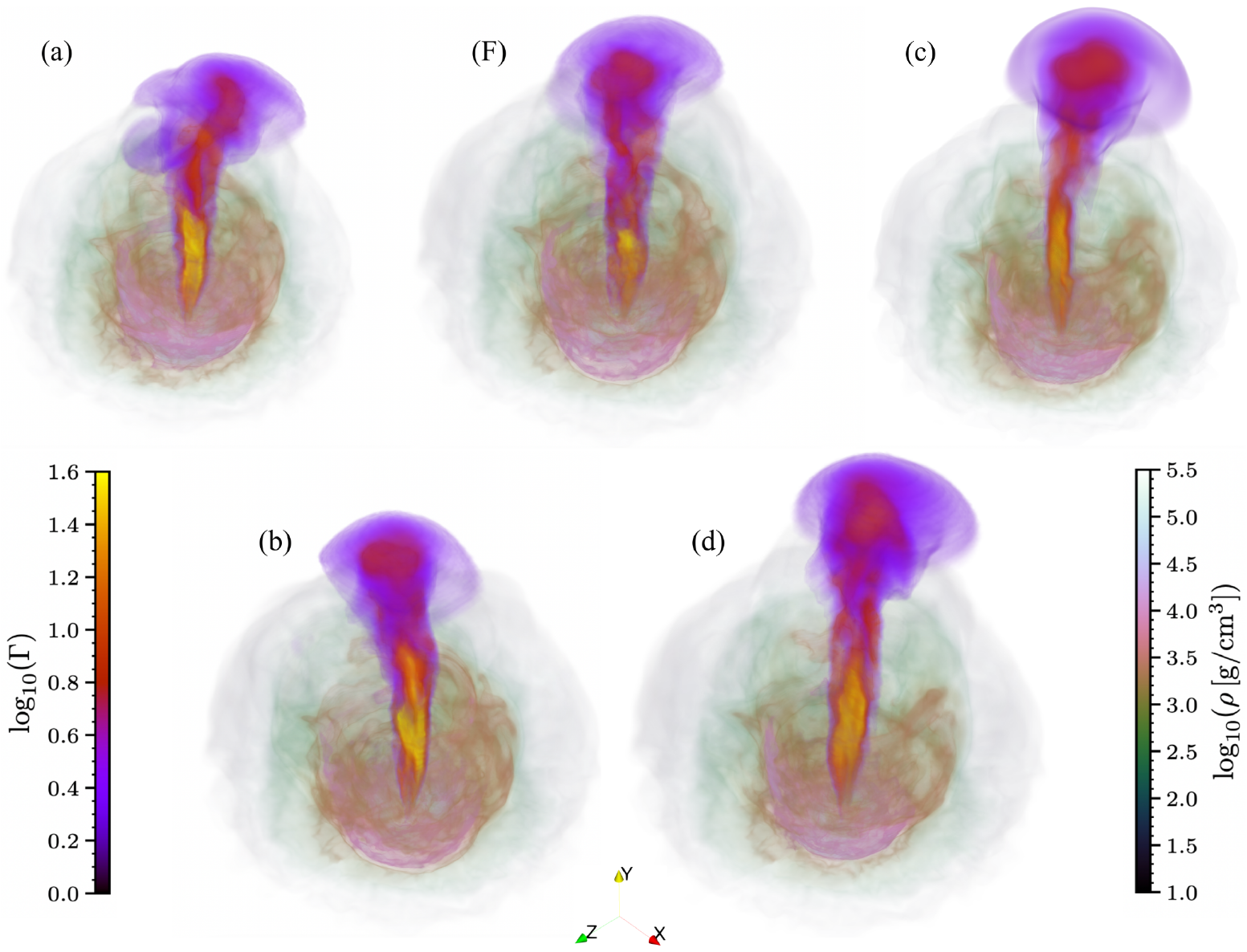}
    \caption{Volume rendering of the jet-environment system configuration at time of jet breakout, in terms of the rest-mass density and Lorentz factor, for different models.  
    Specifically, in models (a), (b), (F), (c), and (d), the configuration shown is at $200$\,ms, $275$\,ms, $275$\,ms, \BUG{$275$\,ms}, and $300$\,ms after launch, respectively (launch time is 385\,ms after merger in all models).
    Only configurations in the north side are shown; we refer to Figure~\ref{fig:fig5} for a fully 3D view of model (F) at 1\,s after launch.}
    \label{fig:fig2}
\end{figure*}

Going back to our model~(F), we note that the jet breakout occurs at $\simeq275$\,ms after launch (red dots in Figure~\ref{fig:fig1}).
\RED{The system's configuration in the north side at that time}, in terms of the 3D logarithmic distributions of rest-mass density and Lorentz factor, is shown at the top center \GREEN{in Figure~\ref{fig:fig2}}.
We see that the BNS merger environment has been excavated by the relativistic jet in all density regions, the latter decreasing outward from $\simeq10^5$ to $\simeq10$\,g/cm$^3$ (notice that variable color transparency is employed in the Figure).
Within such regions, the $\Gamma$ distribution is uniform and smooth near the jet base, while toward the jet front it looks more discontinuous and asymmetrical.
Moreover, the same distribution shows higher values near the jet base: maximum is $\Gamma_\mathrm{max}\simeq48$ below $2\times10^4$\,km radius, including the south side; while $\Gamma_\mathrm{max}\simeq18$ toward the jet front. 
Finally, we note the high collimation of the jet within the dense post-merger environment, and the widening of the jet opening angle upon entry into the rarefied \GREEN{medium.}

\begin{figure*}
    \centering
    \includegraphics[width=.9\columnwidth]{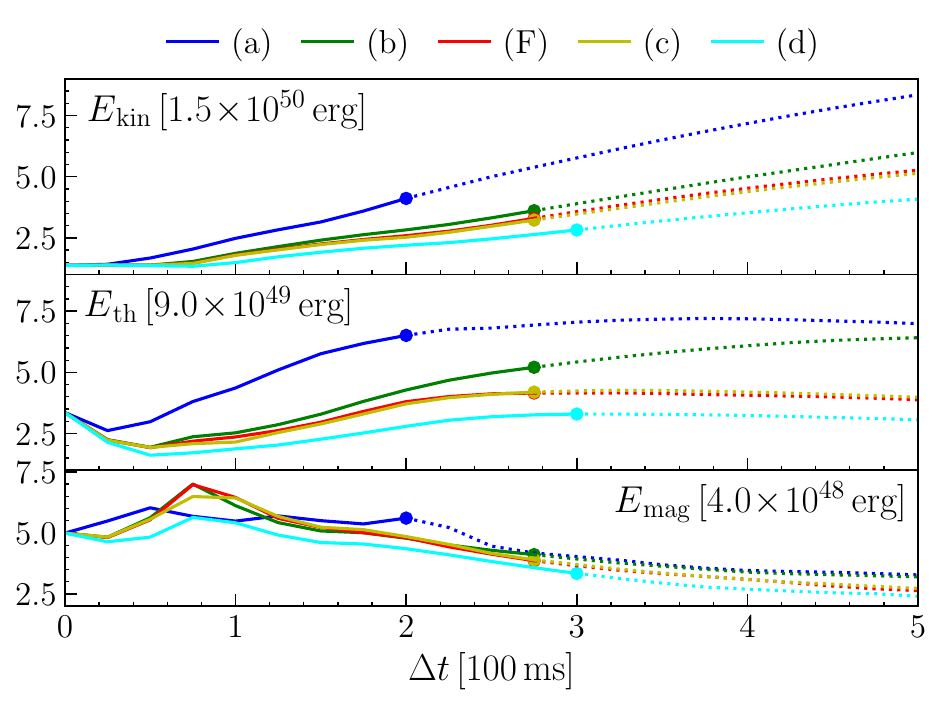}
    \includegraphics[width=.9\columnwidth]{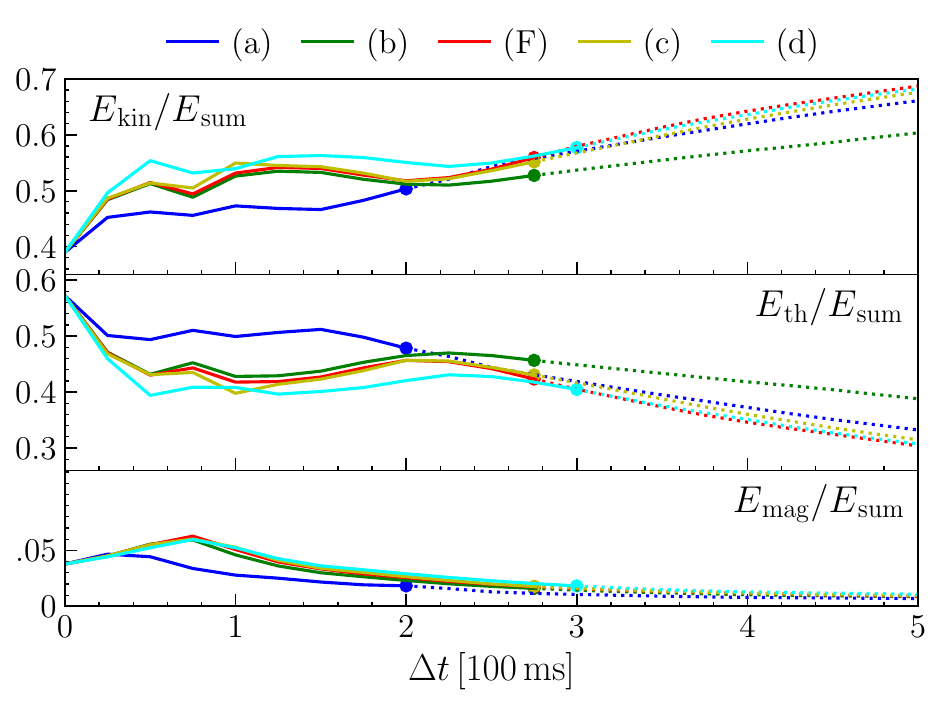}
    \caption{System evolution in terms of kinetic ($E_\mathrm{kin}$), thermal ($E_\mathrm{th}$), and magnetic ($E_\mathrm{mag}$) energy, from launch to 500\,ms later, for different models. The left panel shows $E_\mathrm{kin}$ (top), $E_\mathrm{th}$ (middle), and $E_\mathrm{mag}$ (bottom), while the right panel shows each energy divided by $E_\mathrm{sum}=E_\mathrm{kin}+E_\mathrm{th}+E_\mathrm{mag}$. The solid/dotted colored lines indicate the trends before/after the jet breakout (marked by colored dots).}
    \label{fig:fig3}
\end{figure*}

In Figure~\ref{fig:fig3}, we show the system evolution from launch to 500\,ms later, in terms of kinetic ($E_\mathrm{kin}$), thermal ($E_\mathrm{th}$), and magnetic ($E_\mathrm{mag}$) energy. 
Specifically, these energies are calculated in the lab frame, over the entire computational domain as follows:
\begin{align}
    E_\mathrm{kin} &= \int
    e_\mathrm{kin}\,\mathrm{d}V=\int \rho\Gamma c^2\left(\Gamma-1\right)\mathrm{d}V\,,\label{eq:ekin}\\
    E_\mathrm{th} &= \int e_\mathrm{th}\,\mathrm{d}V=\int[\,\rho\epsilon\Gamma^2c^2+P(\Gamma^2-1)]\mathrm{d}V\,,\label{eq:eth}\\
    E_\mathrm{mag} &= \int e_\mathrm{mag}\,\mathrm{d}V=\int [\,b^2(\Gamma^2-1/2)-(b^0)^2]\mathrm{d}V\,,\label{eq:emag}
\end{align}
where $e_\mathrm{kin}$, $e_\mathrm{th}$, $e_\mathrm{mag}$ are the kinetic, thermal, and magnetic energy densities, respectively, and
\begin{equation}
    b^0 = \dfrac{1}{\sqrt{4\pi}}\Gamma(\vec{v}\cdot\vec{B})/c\,.
\end{equation}
Note that $e_\mathrm{kin}+e_\mathrm{th}+e_\mathrm{mag}$ plus the rest-mass energy density, $\rho\Gamma c^2$, gives us the total energy density in the lab frame, i.e., the time-time component of the energy-momentum tensor in the same frame \citepalias{Pavan2023}. 
In the left panel, in particular, we show the time trends of the above energies, while in the right panel those of each energy divided by $E_\mathrm{sum}=E_\mathrm{kin}+E_\mathrm{th}+E_\mathrm{mag}$. 

At the time of jet launch (in all models in Figure~\ref{fig:fig3}), we have that $E_\mathrm{kin}\simeq2.06\times10^{50}$\,erg and $E_\mathrm{th}\simeq3.02\times10^{50}$\,erg, which correspond to about 39 and 57~per~cent of $E_\mathrm{sum}$, respectively.
On the other hand, $E_\mathrm{mag}\simeq2.00\times10^{49}$\,erg, i.e. $\simeq4$~per~cent of $E_\mathrm{sum}$.
Hence, at launch, pure-HD contributions dominate the energetics of the BNS merger environment as a whole.\footnote{Notice, however, that such \RED{an} environment has magnetic energy density and magnetic pressure higher than the thermal ones beyond $\simeq2500$\,km radius (Figure~7 in \citetalias{Pavan2023}). On the other hand, thermal contributions dominate by far at lower radii, resulting in $E_\mathrm{th}\gg E_\mathrm{mag}$ when integrating over the entire domain.}

\RED{In model (F)}, the kinetic energy remains \RED{essentially} constant until $\simeq\!80$\,ms after launch (with less than $7$~per~cent of variation).
In such a time window, in fact, the jet \RED{struggles} to pierce the higher-density regions of the surrounding environment, with the result that much of the injected energy is deposited there instead of being converted into bulk acceleration.
At very early stages, in addition, a significant drop in thermal energy ($\simeq50$~per~cent) and a minor drop ($\simeq15$~per~cent) in magnetic energy occur.
Both are due to the falling back of dense, hot and magnetized material below the excision radius, driven by the gravitational pull from the central object of mass $M_0$. Starting from $\simeq\!40$\,ms after launch, then, a remarkable change in the time trend of such energies is observed.
Due to the first successes of the jet in drilling through the environment, a significant amount of injected energy is brought into the system, increasing $E_\mathrm{th}$ and $E_\mathrm{mag}$.
Notice the relative steep increase in the magnetic energy from $\simeq40$\,ms to $\simeq80$\,ms after launch. 
This occurs because at such early times the jet drilling and associated energy conversion process are not yet well established, thus the injected magnetic energy accumulates within the system itself.
Thereafter, $E_\mathrm{mag}$ decreases rapidly, being converted into bulk acceleration and thermal energy, the latter being in turn converted to $E_\mathrm{kin}$ via adiabatic expansion.\footnote{\RED{We note that magnetic energy may also be dissipated via resistive processes; however, in our simulations (where resistivity is of numerical origin) such processes are subdominant, as demonstrated in Appendix~\ref{app:res_stud}.}}

By looking at the right panel in the same Figure, we note a nearly anti-symmetric, non-monotonic trend of $E_\mathrm{kin}/E_\mathrm{sum}$ with respect to $E_\mathrm{th}/E_\mathrm{sum}$ before the jet breakout.
\GREEN{In particular,} the non-monotonic pattern results from the turbulent interaction of the jet with the realistic BNS merger environment, which causes repeated modifications of the energy partitioning in the whole system.
After the jet breakout, on the other hand, \GREEN{$E_\mathrm{kin}/E_\mathrm{sum}$} monotonically increases over $E_\mathrm{th}/E_\mathrm{sum}$ due to acceleration into the rarefied medium.

Thermal and magnetic energy density distributions at breakout, in the $xy$-plane (north side only), are shown at the top center in Figure~\ref{fig:fig4}, in the left and right polyptychs, respectively.
In each panel, the white horizontal dashed line indicates the distance reached by the jet front at that time (consistent with values of $r_\mathrm{j}$ in Figure~\ref{fig:fig1}).
We observe that $e_\mathrm{th}$ is higher near the injection site and along the channel excavated by the jet. 
\GREEN{On the other hand}, $e_\mathrm{mag}$ peaks only around the former.  
This means that \RED{in model (F)}, at breakout, the magnetic energy in the jet has largely been converted into kinetic and thermal \GREEN{forms.}
\RED{Notably}, the high collimation of the jet up to the breakout radius (\GREEN{$\simeq4.2\times10^4$\,km}) 
is not due to confinement by the jet magnetic field, but is a consequence of the wider-angle cocoon surrounding the jet itself, in which a considerable amount of energy is \GREEN{stored}.

\GREEN{The} configuration reached at 1\,s after launch is shown in Figure~\ref{fig:fig5}.
Specifically, the upper \RED{central panel displays} the distributions of rest-mass density and Lorentz factor in the \GREEN{$xy$-plane}, while the lower panel \RED{provides} a full 3D view of the same \GREEN{distributions} (\RED{notice} that variable color transparency is \RED{consistently used in these panels}).
\RED{We observe} a low degree of north-south asymmetry \RED{in model (F)}, with $\Gamma_\mathrm{max}\simeq49.2$ and $\simeq58.6$ in the north and south sides, respectively.

\RED{We refer the reader to Section~\ref{sec:angular3s} for the properties of the jet at 3 s after launch, corresponding to much larger spatial scales than those analyzed so far.}

\subsection{Luminosity and decay time scale}
\label{sec:L_tau}

The impact of luminosity on the propagation of GRB jets has been investigated both in analytical/semi-analytical models \citep[e.g.,][]{Harrison2018,Salafia2020,Hamidani2021,Gottlieb2022b,Garcia-Garcia2024} and in numerical simulations reproducing jet injection and propagation for different values of initial jet luminosity and decay time scale \citep[e.g.,][]{Duffell2018,Gottlieb2021,Murguia2021,Nativi2021,Nativi2022,Urrutia2021,Urrutia2023}.
In these investigations, it has been shown that as the initial jet luminosity increases, the jet front propagates faster into the same BNS merger environment, shortening the breakout time.
On the other hand, when such \GREEN{a} luminosity decreases, the propagation results to be slower, and the breakout itself may not occur (\GREEN{`choked jet'}).
Finally, in the case of a jet injection luminosity decaying over time, the decay time scale affects the dynamics of propagation and the eventual success of the breakout.

In this Subsection, we investigate the impact of luminosity and its decay time scale on the propagation of GRB jets in a realistic BNS merger environment.
Specifically, we discuss and compare with our previous fiducial model (F) the results of the following simulations:

$\bullet$ model (a), with setup similar to model (F) but initial jet luminosity increased by a factor $5/3$, i.e. \BUG{$L_\mathrm{j}=8.95\times10^{51}$\,erg/s};

$\bullet$ model (b), with setup similar to model (F) but decay time scale of luminosity increased by a factor $5/3$, i.e. $\tau_\mathrm{d}=500$\,ms.

In both models, we perform the launch of the jet at the same time as in model (F), i.e. 385\,ms after merger, into the same realistic environment, \RED{using the same prescription described in Subsection~\ref{sec:jet}.}
However, given the increased $L_\mathrm{j}$ and $\tau_\mathrm{d}$ in models (a) and (b), respectively, the energy injected at $r=\rexc$ over the total simulation time is a factor $\approx5/3$ larger than in model (F).
In addition, we keep the same $\Sigma_\mathrm{j}$ as in model~(F). 
This allows us to probe the impact of luminosity and its decay time scale under the same relative contribution of the jet magnetic \GREEN{field}.
To achieve this in model~(a), we increase the maximum toroidal magnetic field of the jet \RED{to $B^{\Phi}_\mathrm{j,m}=1.56\times10^{13}$\,G}.
Refer to Table~\ref{tab:1}, for the detailed configuration of each model.
\begin{figure*}
    \centering
    \includegraphics[width=0.99\columnwidth]{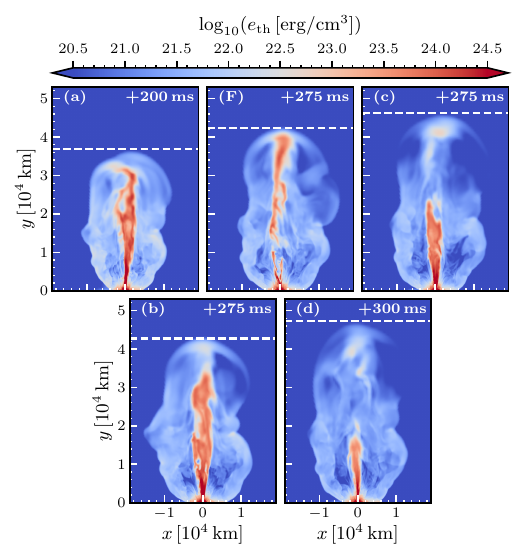}
    \includegraphics[width=0.99\columnwidth]{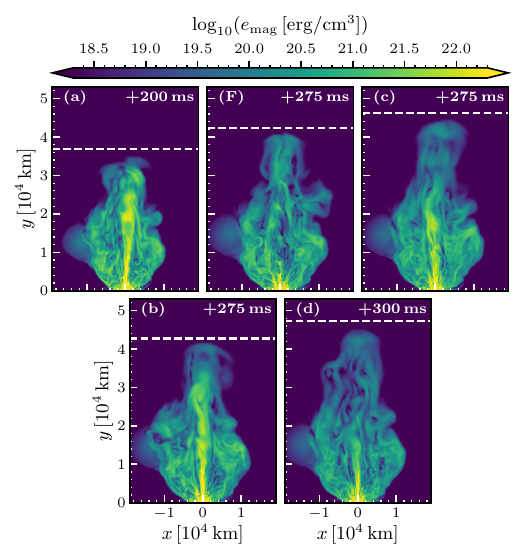}
    \caption{Thermal (left polyptych) and magnetic (right polyptych) energy densities at the time of jet breakout, for different models (north side only).
    Within each panel, the specific model, time of jet breakout (relative to launch), and distance along $y$-axis reached by the jet front are shown \GREEN{in white}.}
    \label{fig:fig4}
\end{figure*}

In Figure~\ref{fig:fig1}, the time trends of $r_\mathrm{j}$ in models (a) and (b) are indicated with blue and green lines, respectively. 
Compared to model~(F), we note that $r_\mathrm{j}$ reaches much larger values in model (a), both in the north and south sides, and in all 500\,ms after the jet launch.
Moreover, the jet breakout occurs $\simeq\!75$\,ms earlier, i.e. $\simeq200$\,ms after launch (blue dots in the Figure).
Thus, the increase in $L_\mathrm{j}$ adopted in model~(a) improves jet drilling in the realistic BNS merger environment, making jet propagation faster.
This is in agreement with the results of previous investigations mentioned above, despite the use of an analytical environment in most of them.
Comparing models (F) and~(b), on the other hand, we do not see significant differences both in the north and south sides, and both before and after the jet breakout ($\simeq275$\,ms after launch in both models).
\RED{Thus, the increase in $\tau_\mathrm{d}$ has a minor impact on jet drilling, particularly on the jet front kinematics (but see the discussion below regarding the overall jet energetics).}
\begin{figure*}
    \centering
    \includegraphics[width=1.76\columnwidth]{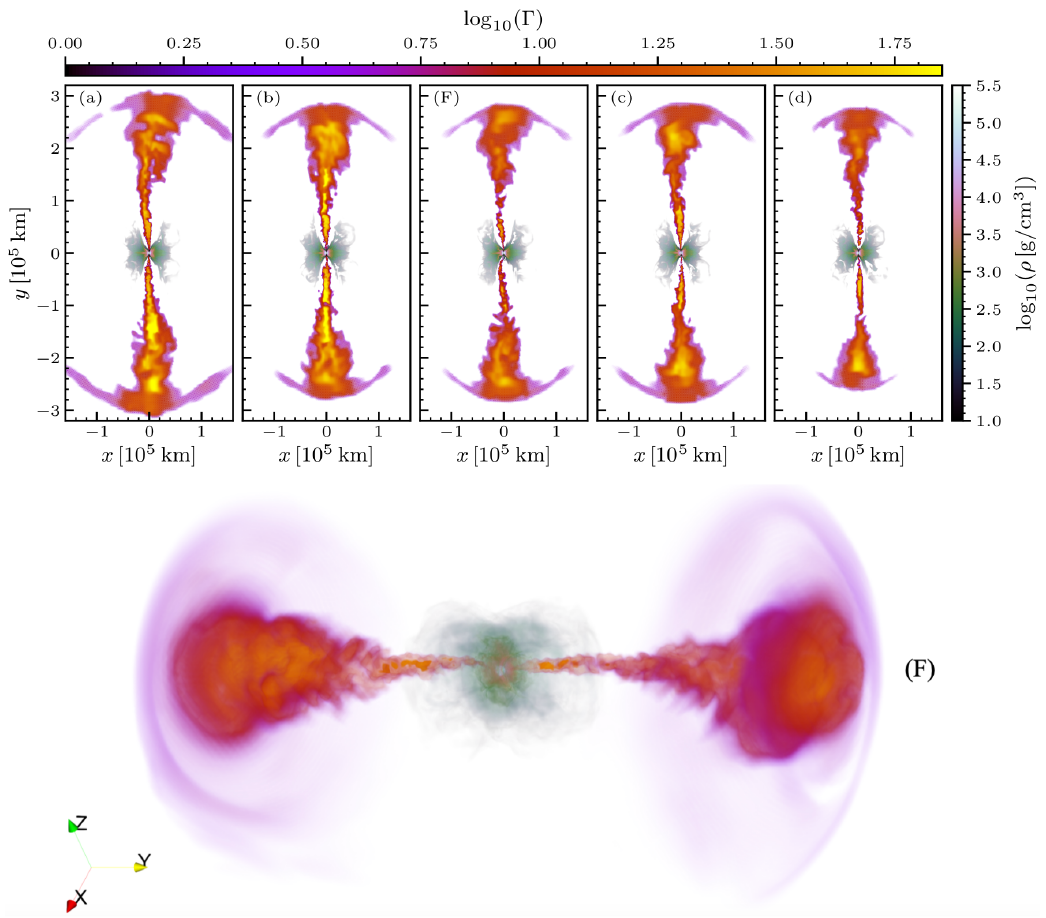}
    \caption{Jet-environment system configuration at 1\,s after launch, in terms of rest-mass density and Lorentz factor, for different models. The upper panels show the north-south distributions in the $xy$-plane, for different models, while the lower panel the fully 3D view of the same distributions in the fiducial \GREEN{model (F)}.}
    \label{fig:fig5}
\end{figure*}

Comparing with analytical modeling by \cite{Gottlieb2022b}, for $\alpha=4$ (refer to Subsection~\ref{sec:fiducial}), we have that (Eqs.~20,25 of the cited work, adapted for the present discussion):
\begin{equation}\label{eq:rj_trend}
    r_\mathrm{j}(\Tilde{t})=\rexc
    \begin{cases}
        1+\dfrac{\Theta_\mathrm{j}^{-2/3}-1}{\Tilde{t}_\mathrm{coll}-1}\left(\Tilde{t}-1\right)\quad\quad&\Tilde{t}<\Tilde{t}_\mathrm{coll}\\
        \Theta_\mathrm{j}^{-2/3}\left(\dfrac{\Tilde{t}-1}{\Tilde{t}_\mathrm{coll}-1}\right)^3\left(\dfrac{\Tilde{t}}{\Tilde{t}_\mathrm{coll}}\right)^{-1}\quad\quad&\Tilde{t}_\mathrm{coll}<\Tilde{t}
    \end{cases}\quad,
\end{equation}
where $\Tilde{t}=t/(t_\mathrm{c}+\tau_\mathrm{j})$ and  $\Tilde{t}_\mathrm{coll}\approx3/\eta_\mathrm{coll}+1$, being 
\begin{equation}\label{eq:eta_coll}
    \eta_\mathrm{coll} \approx 0.4\left(\dfrac{\eta_\mathrm{crit}}{N_\mathrm{crit}}\right)^{2/3}\Theta_\mathrm{j}^{-4/9}\,.
\end{equation}
In addition, according to Eq.~\ref{eq:eta_crit}, we have that $\eta_\mathrm{crit}\propto L_\mathrm{j}^{1/2}$ for given environment configuration at the time of launch ($t_\mathrm{c}+\tau_\mathrm{j}$), and initial jet half-opening angle ($\Theta_\mathrm{j}$). 
Including this in the above equations, we have that as $L_\mathrm{j}$ increases, $\Tilde{t}_\mathrm{coll}$ decreases, resulting in higher values of $r_\mathrm{j}$ at any time after the launch.
Moreover, $r_\mathrm{j}$ grows faster for $\Tilde{t}_\mathrm{coll}<\Tilde{t}$ (Eq.~\ref{eq:rj_trend}), and then such \RED{a} trend starts earlier as $L_\mathrm{j}$ increases.
All this is actually observed in Figure~\ref{fig:fig1} when comparing models (F) and (a).
In particular, in the former, a faster growth in time of $r_\mathrm{j}$ is observed starting from $\simeq80$\,ms after launch, while in the latter, with increased $L_\mathrm{j}$, as early as $\simeq20$\,ms after launch. 
Notably, we observe that in both models \RED{the} faster growth \RED{of} $r_\mathrm{j}$ starts upon reaching full collimation of the jet at its base, due to the formation of a pressured cocoon around it.
Indeed, this is exactly the process that is expected to occur at~$\Tilde{t}_\mathrm{coll}$ according to the aforementioned modeling.
Using the corresponding analytical expression, we estimate \BUG{$\Tilde{t}_\mathrm{coll}\simeq1.204$} in model~(F), i.e. full collimation of the jet is expected to occur at \BUG{$\simeq78.5$\,ms} after launch.
This is highly consistent with our numerical result (differences less than time interval between two consecutive simulation outputs), despite the realistic environment taken into account in our model.
This is because at the jet base, where the above collimation process takes place, interaction with such an environment is rather weak, reducing the impact of the latter on evolution. 

The north side system configuration at the time of jet breakout in models (a) and (b) is shown at the top left and bottom left in Figure~\ref{fig:fig2}, respectively.
In both models, we note a high-$\Gamma$ region extending from the base of the jet up to its middle, which is more elongated \RED{and heightened compared to} the corresponding region in model (F)\RED{, indicating a more stable and} efficient conversion of the injected energy into kinetic \GREEN{form}.

Focusing then on energetics, in the left panel in Figure~\ref{fig:fig3} the time trends of $E_\mathrm{kin}$, $E_\mathrm{th}$, and $E_\mathrm{mag}$ in models (a) and (b) are indicated with blue and green lines, respectively.
We note that $E_\mathrm{kin}$ reaches higher values in model (a) than in models (b) and (F), in all 500\,ms after the jet launch \RED{(consistent with enhanced jet front kinematics).}
The same occurs for $E_\mathrm{th}$, due to the larger amount of injected energy in model (a) from the very beginning of the simulation.
On the other hand, we note very similar trends of $E_\mathrm{kin}$ and $E_\mathrm{th}$ at early times after launch in models (b) and (F), \RED{while at later times, the longer $\tau_\mathrm{d}$} adopted in the former model results in a larger increase in $E_\mathrm{kin}$ and $E_\mathrm{th}$ over time\RED{, with $E_\mathrm{th}$ reaching values comparable to those} in model (a) at the end of the 500\,ms time \GREEN{window}.

The energy conversion in models (a) and (b) can be evaluated by looking at the right panel in Figure~\ref{fig:fig3}, blue and green lines, respectively.
Here, we note similar features as those discussed in model~(F) \RED{(see Subsection~\ref{sec:fiducial})}. 
However, the increased $L_\mathrm{j}$ adopted in model (a) results in larger values of $E_\mathrm{th}/E_\mathrm{sum}$ before the jet breakout compared to model (F).
As we show below, such \RED{a larger fraction} of thermal energy leads the jet to accelerate toward higher Lorentz factors during the subsequent free expansion phase.
Higher $E_\mathrm{th}/E_\mathrm{sum}$ is observed in model (b) as well, but this occurs at longer times after launch than in model (a), \RED{as a result of the slower decay of the jet power over time.}

The above discussion is enriched when looking at the $e_\mathrm{th}$ and $e_\mathrm{mag}$ distributions in models (a) and (b) in Figure~\ref{fig:fig4}, at the top left and bottom left in each polyptyich, respectively. 
At breakout, the increased $L_\mathrm{j}$ and $\tau_\mathrm{d}$ in models (a) and (b), respectively, results in more continuous and heightened distributions than in model (F).
\RED{In particular,} $e_\mathrm{mag}$ peaks both at the jet base and along the jet propagation axis, not only around the former as in model (F).
This means that the magnetic energy has not all been spent on jet \GREEN{drilling,} but it will contribute to accelerate the jet during free expansion.

\RED{The Lorentz factor distributions reached at 1\,s after launch in mo\-dels (a) and (b) are shown} in the first and second panels from the left in Figure~\ref{fig:fig5}, respectively.
In both models, the jet \RED{attains much higher values of $\Gamma$} than in \GREEN{model (F)}.
Specifically, when taking into account the full jet distribution, in the north side we obtain $\Gamma_\mathrm{max}\simeq81.6$ and $\simeq125.9$ in models (a) and (b), respectively, while in the south side $\simeq93.7$ and $\simeq98.9$, respectively (cfr.~Subsection~\ref{sec:fiducial}).

In Section~\ref{sec:angular3s}, the jet configuration at 3\,s after launch in models~(a) and (b), respectively, will be discussed and compared with that in the other models, to provide final insights into the impact of luminosity and decay time scale.

\subsection{Magnetic field contribution}
\label{sec:magn_content}

Magnetic fields play a central role in BNS mergers and their connection with GRBs (see \citealt{Ciolfi2020b} for a review).
Notably, the winding of magnetic fields threading a rotating BH surrounded by an accretion disk is to date considered one of the leading mechanisms for launching a GRB jet in a BNS merger aftermath \citep[see][and refs. therein]{Blandford1977,Thorne1986,Lee2000}.
Therefore, magnetic fields are very likely to be involved in the subsequent jet propagation through the surrounding BNS merger environment.
Their impact on the propagation process has been investigated through (G)RMHD simulations, either within the ideal MHD framework (e.g., \citealt{Bromberg2018,Geng2019,Kathirgamaraju2019,Gottlieb2020,Nathanail2020,Nathanail2021,Gottlieb2022a}; \citetalias{Pavan2023}; \citealt{Garcia-Garcia2023}) or taking into account physical resistivity \citep{Mattia2024}. 

In \cite{Gottlieb2020}, in particular, a detailed comparison between the propagation of weakly magnetized jets ($\sigma\sim0.01-0.1$; Eq.~\ref{eq:sigma_small}) and that of purely hydrodynamic jets has been carried out, showing that in the former case instabilities at the jet-environment interface are significantly reduced, resulting in more efficient drilling and, consequently, a jet front propagation speed two to three times faster.
\GREEN{Conversely}, in a parallel analysis of core-collapse GRB jets, \cite{Bromberg2016} have observed that in the case of dominant magnetic fields (i.e.~$\sigma\!\gg\!1$), the jet dynamics becomes prone to kink instabilities, developed both internally to the jet and along the aforementioned interface, which induce magnetic field dissipation and global deformation of the jet angular structure over a time scale (measured in the lab frame) 
\begin{equation}\label{eq:kink}
    t_\mathrm{kink}\simeq\dfrac{2\pi R_0\Gamma_\mathrm{j}}{c}\dfrac{|b^\mathrm{Pol}|}{|b^\mathrm{Tor}|}\sqrt{1+\dfrac{1}{\sigma}}\,,
\end{equation}
where $b^\mathrm{Pol}$ and $b^\mathrm{Tor}$ are, respectively, the poloidal and toroidal magnetic field strengths measured in the proper frame, and $R_0$ is the size of the jet core dominated by $b^\mathrm{Pol}$.

In this Subsection, we investigate the impact of magnetic fields on jet propagation through a realistic magnetized BNS merger environment.
Specifically, we discuss and compare with our fiducial model (F) the results of the following simulations:

$\bullet$ model (c), with setup similar to model (F) but maximum toroidal magnetic field increased to \BUG{$B^{\Phi}_\mathrm{j,m}=1.38\times10^{13}$\,G};\footnote{Across our models, the adopted values of $B^{\Phi}_\mathrm{j,m}$ (at $\rexc=380$\,km; Table~\ref{tab:1}) would correspond to magnetic field strengths at a radius of 10\,km (i.e.~close the poles of the central black hole) in the range $(0.9-2.3)\times 10^{16}$\,G. These values are consistent with those required to launch a GRB-like incipient jet in the aftermath of a BNS merger, as observed in GRMHD simulations (e.g., \citealt{Ruiz2016,Gottlieb2022a}).}

$\bullet$ model (d), with setup similar to model (F) but with the same relative contribution of magnetic luminosity as in model (c), i.e.~\BUG{$\Sigma_\mathrm{j}=1.56\times10^{-2}$} (see also below).

In both models, \RED{in particular,} we perform the launch of the jet at the same time as in model (F), into the same realistic environment (similarly to models (a) and (b) in Subsection~\ref{sec:L_tau}).
\RED{Nevertheless, $\sigma\!\in\![0.03,1.52]$ in models (a), (b), and (F), whereas $\sigma\!\in\![0.04,4.89]$ in models (c) and (d).}
In the latter model, moreover, we adopt the same $B^{\Phi}_\mathrm{j,m}$ as in model (F), lowering $L_\mathrm{j}$ \RED{to keep} the same $\Sigma_\mathrm{j}$ as in model (c).
This allows us to probe the impact of increased jet magnetization at a lower initial jet luminosity than in model (F).
Refer to Table~\ref{tab:1}, for the detailed configuration of each model.

In Figure~\ref{fig:fig1}, the time trends of $r_\mathrm{j}$ in models (c) and (d) are indicated with yellow and cyan lines, respectively. 
Compared to model~(F), we see that $r_\mathrm{j}$ reaches \BUG{similar} values in model~(c), both in the north and south sides, \BUG{with the jet breakout that occurs at $\simeq275$\,ms after launch}.
\GREEN{The} average advancement speed at the jet front (calculated as the temporal variation of $r_\mathrm{j}$ between two consecutive simulation output times) \BUG{is comparable between these models, while $\simeq\!2.5$ times higher} than in the model with $B^{\Phi}_\mathrm{j,m}\!=\!0$.
\GREEN{Such an increase is due to both the higher initial jet luminosity and the presence of jet magnetic fields, the latter being consistent with \citet [][see above discussion]{Gottlieb2020}.
Notably, the increase in average advancement speed due to a higher initial jet luminosity alone, as estimated between models (F) and (a), is by a factor $\simeq\!1.6-1.7$.}

Comparing models (d) and (F) in Figure~\ref{fig:fig1}, then, we note similar time trends in the north side, while in the south side, $r_\mathrm{j}$ reaches lower values in model~(d).
In such \RED{a} model, moreover, in the north side the jet breakout occurs $\simeq25$\,ms later than in model (F), i.e. $\simeq300$\,ms after the jet launch, while in the south side it is further delayed by $\simeq25$\,ms.
This is due the lower $L_\mathrm{j}$ adopted in model (d), which makes it harder for the jet to drill into the same BNS merger \GREEN{environment}.
The more difficult drilling along with the realistic anisotropies of the environment itself account for the observed north-south asymmetries in the jet propagation \GREEN{trends}.\footnote{Some north-south asymmetries can develop also for isotropic magnetized environment, as discussed in Appendix~\ref{app:improvements}, even though they tend to be less pronounced than for realistic BNS merger environments.}

\begin{figure*}
    \centering    
    \includegraphics[width=1.2\columnwidth]{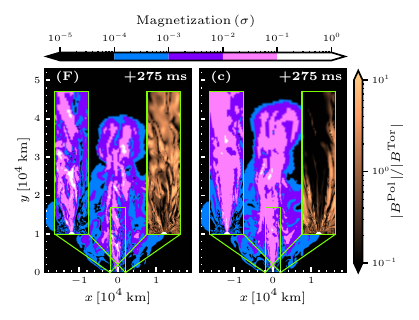}
    \caption{Magnetization and lab-frame poloidal-to-toroidal magnetic field ratio in the $xy$-plane, north side, at the time of jet breakout in models (F) and (c) (left and right panels, respectively). See discussion in the text.}
    \label{fig:fig6}
\end{figure*}
The north-side configuration at the time of jet breakout in models (c) and (d) is shown at the top right and bottom right in Figure~\ref{fig:fig2}, respectively.
\RED{Similar to models (a) and (b)}, we note the presence of a high Lorentz factor region extending from the base of the jet up to its middle, which is more elongated relative to the total jet size than the corresponding region in model~(F).
\RED{Such a pattern is also evident in the distributions shown in Figures~\ref{fig:fig4} and \ref{fig:fig5}, and is associated with better preservation of the initial jet toroidal magnetic field}.

Preservation of the initial jet \RED{toroidal magnetic field is highlighted} in Figure~\ref{fig:fig6}, in which distributions of magnetization and lab-frame poloidal-to-toroidal magnetic field ratio in the $xy$-plane, in the north side, are reported at the time of jet breakout in models (F) and (c), \GREEN{respectively}.
We notice that in model (F), \GREEN{there} is a less continuous and on average weaker magnetization within the jet cone, associated with a predominant poloidal component of the magnetic field (right inset in the left panel). 
In model (c), conversely, a more continuous and on average higher magnetization is associated with a predominant toroidal magnetic field within the jet cone (right inset in the right panel).
We emphasize that the higher poloidal-to-toroidal magnetic field ratio observed \GREEN{in model (F)} is not a result of kink instabilities, such as those reported by \cite{Bromberg2016}.
If this were the case, at fixed $R_0$, $\Gamma_\mathrm{j}$, and $|b^\mathrm{Pol}|/|b^\mathrm{Tor}|$ in the~same jet propagation medium (\RED{refer to} Table~\ref{tab:1}), we would expect to observe an increase of kink instabilities and associated dissipation of the toroidal magnetic field as the jet magnetization increases \RED{(Eq.~\ref{eq:kink}; see further discussion in the next Section).}

We refer to Section~\ref{sec:angular3s}, for the jet configuration at 3\,s after launch in models (c) and (d), respectively, and further insights into the impact of magnetic fields on jet angular structure and energetics.

\section{Choking condition and launch time}
\label{sec:choking_tc}

\GREEN{The impact} of a different launch time on jet propagation within BNS merger environments has been already explored via numerical simulations.
In \cite{Murguia2017,Murguia2021}, in particular, the authors have simulated jet propagation within baryonic winds expelled for a time duration $t_\mathrm{w}$ ranging from 0.3\,s to 1.0\,s prior to jet launch ($t_\mathrm{w}$ being related to the time it takes for a hypothetical NS remnant to collapse into BH).
The jet launch has been initiated manually at the end of the wind phase, using a parametrized prescription accounting for a variable launch duration (namely $t_\mathrm{j}$, ranging from 0.5\,s to 1.0\,s).
Analysis of the simulation results has shown that for a longer $t_\mathrm{w}$ (in particular a longer $t_\mathrm{w}/t_\mathrm{j}$) the interaction of the jet with the baryonic wind is enhanced, thus energy deposition into the latter is increased.
In this case, if the jet velocity at the front becomes lower than that of the upstream material, or if launch terminates before the jet reaches the outer boundary of the wind, the breakout does not occur, i.e.~the jet is choked.

Similar results have been obtained in simulations by \cite{Geng2019}, which have also taken into account magnetic fields.
Specifically, using a magnetized jet prescription similar to the one described in Subsection~\ref{sec:jet}, 
the authors have simulated jet propagation within (analytical) dynamical ejecta from BNS merger, which expand for a time duration $\Delta t_\mathrm{jet}$ ranging from 0.01\,s to 1\,s prior to jet launch.
Analysis of the simulation results has shown that for a shorter $\Delta t_\mathrm{jet}$ (e.g. 0.1\,s or less) the breakout is easier and occurs earlier in time, due to weaker interaction of the jet with the merger ejecta.
This also leads to a final system angular structure closer to a `top-hat' jet, that is, an ultra-relativistic uniform conical outflow.
Conversely, a longer $\Delta t_\mathrm{jet}$ results in a later breakout time, as well as a more structured final outflow, characterized by a mildly relativistic wide-angle region surrounding a narrow ultra-relativistic core.
\begin{figure*}
    \centering    
    \includegraphics[width=1.75\columnwidth]{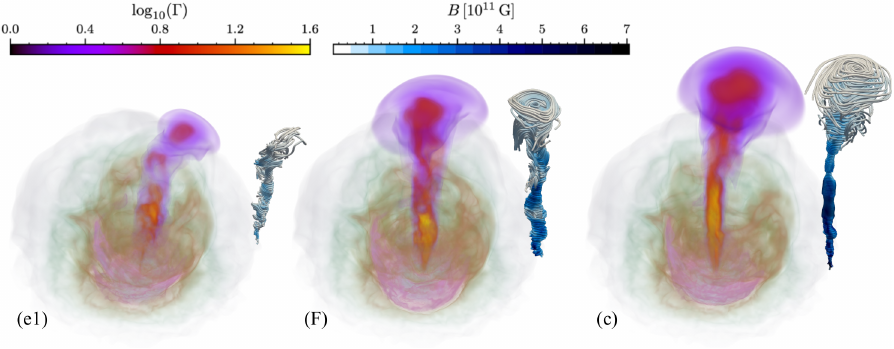}
    \caption{Volume rendering of the rest-mass density and Lorentz factor distributions, and jet magnetic field configuration at \BUG{275\,ms} after launch, north side, in models (e1), (F), and (c), respectively. The same colorbars used in Figure~\ref{fig:fig2} are adopted here. Moreover, the jet magnetic field is represented through streamlines with shades of blue varying according to the local field strength in the lab frame ($B$).}
    \label{fig:fig7}
\end{figure*}

The above results have been derived in 2D (or 2.5D) simulations, which, however, do not reproduce instabilities that notoriously affect jet propagation \citep[e.g.][see also previous Section]{Gottlieb2021}.
Moreover, in these simulations, simplified profiles for the jet propagation environment have been employed.\footnote{In \cite{Murguia2017,Murguia2021}, rest-mass density profiles have been derived by averaging the azimuthal profiles obtained in the simulations by \cite{Perego2014,Siegel2014,Janiuk2019}, and a constant velocity has been adopted.}

In \cite{Pavan2021}, we presented 3D RHD simulations evaluating the impact of jet launch time on propagation within realistic BNS merger environments.
Specifically, in our simulations the same GRB jet was manually injected into two different environments imported, respectively, at $\simeq100$\,ms and $\simeq200$\,ms after merger from the reference GR simulation.
The resulting physical evolution was consistent with the above findings in terms of variations in the breakout time, and final system angular structure.
In addition, our simulations showed that anisotropies characteristic of realistic BNS merger environments can significantly reduce the jet breakout time compared to analytical isotropic media.

Based on these considerations, here we provide a more advanced analysis of the impact of jet launch time on propagation within realistic BNS merger environments, taking into account magnetic fields both in the jet and the surrounding ambient medium.
Specifically, we discuss and compare with models presented in the previous Section the results of the following simulations:

$\bullet$ model (e1), with setup similar to fiducial model (F) but \GREEN{$L_\mathrm{j}$ and $B^{\Phi}_\mathrm{j,m}$ lowered to, respectively,} \BUG{$3.69\times10^{51}$\,erg/s and $0.62\times10^{13}$\,G};

$\bullet$ model (e2), with setup similar to model (e1) but jet launch time reduced by 200\,ms, that is, equal to 185\,ms after merger.

In both models (e1) and (e2), in particular, \BUG{$\Sigma_\mathrm{j}\!=\!0.46\times10^{-2}$}. \GREEN{Moreover, in model (e2), the jet launch occurs after running the 30\,ms collapse phase right at the import time (see introduction of Section~\ref{sec:5jets})}.

In Figure~\ref{fig:fig7}, we show the 3D rest-mass density and Lorentz factor distributions at \RED{275\,ms} after launch (same colorbars as in Figure~\ref{fig:fig2}), and the jet magnetic field configuration at the same time, in the north side, in models (e1), (F), and (c), \GREEN{respectively}.
In all three models, the magnetic field is represented through streamlines (shown on the \RED{right} side of the above distributions for a cleaner view) with shades of blue varying according to the local field strength in the lab frame.
Specifically, such streamlines have been computed in regions where $\Gamma\gtrsim1.75$ to differentiate the ultra-relativistic jet from its mildly relativistic surroundings (both magnetized).

\RED{Comparing models (F) and (c),} we notice a sharper and thus more stable jet-environment interface \RED{in the latter model, associated with better preservation of the initial jet's helical magnetic field. This indicates that, for fixed initial jet luminosity, such a field} is better preserved when instabilities at the \RED{interface (e.g., Rayleigh-Taylor and Kelvin-Helmholtz instabilities; see} \citealt{Gottlieb2021,Pavan2021}) are suppressed by a stronger jet magnetization.

Focusing \GREEN{on} model (e1), we see that at \RED{275\,ms} after launch the jet \RED{is still facing difficulties in emerging from} the realistic BNS merger environment.
This is \RED{due to the relatively low initial jet's luminosity and magnetization, which make} the drilling so difficult that most of the \RED{jet's} energy is deposited into the surroundings instead of being converted into kinetic form (cfr.~Lorentz factor achieved at the jet front in models (e1), (F) and (c)). 
In addition, the collimation and magnetic field configuration \RED{of the jet} are significantly compromised at \RED{275\,ms} after launch.
As we will see in the next Section, the same jet will end up lacking a high Lorentz factor core in the emerging angular distribution (which is instead present for the \GREEN{other models)}.

\begin{figure}
    \centering    
    \includegraphics[width=0.65\columnwidth]{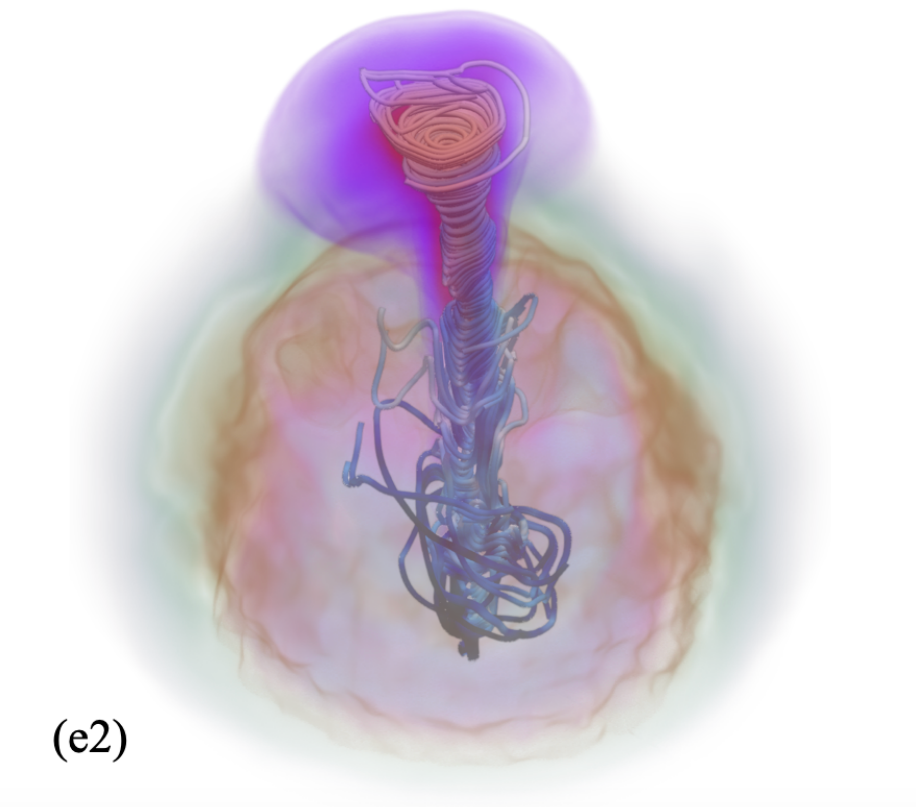}
    \caption{Same as in Figure~\ref{fig:fig7} but at 100\,ms after launch in model (e2). The spatial scale is increased by a factor of $\simeq2.7$ compared to Figure~\ref{fig:fig7} to enhance system visualization. See discussion in the text.}
    \label{fig:fig8}
\end{figure}

Evolution in model (e1) is very different from that in model~(e2), in which the same \RED{low-luminosity and} low-magnetization jet is laun\-ched 200\,ms earlier, namely, at 185\,ms after merger.
At this time, the realistic environment extends along the jet injection axis up to $\simeq3500$\,km instead of \GREEN{$\simeq15000$\,km}, and the mass enclosed within the initial jet opening angle (namely $10^\circ$ from the axis) above the radius $\rexc=380$\,km, $\simeq0.83\times10^{-3}\,M_{\sun}$ both in the north and south sides, is about half compared to 385\,ms after merger ($\simeq1.66\times10^{-3}\,M_{\sun}$).
The reduced spatial extension and mass of the environment encountered by the jet in model (e2) results in a rather short breakout time, only 100\,ms after launch. The corresponding configuration is shown in \GREEN{Figure~\ref{fig:fig8}}.
Notably, the emerging outflow exhibits high collimation and Lorentz factor at the front, associated with toroidally dominated magnetic field configuration (similar to that observed in \GREEN{model (c)}), which is a completely different outcome compared to model (e1).

We conclude that a reduced jet launch time with respect to merger (in our case by 200\,ms, i.e.~a factor $\simeq\!2$) has a major impact on the propagation process. 
Specifically, it shortens the duration of the jet-environment interaction (due to the reduced mass and extension of the environment) and strongly reduces jet energy losses, increasing the chances of a successful \GREEN{breakout}. 

\section{Jet configuration at 3 s after launch}\label{sec:angular3s}

\GREEN{In this Section}, we examine the jet configuration at the end of our simulations (i.e., 3\,s after launch), \RED{probing} how injection parameters influence the jet head's energetics and angular structure, taking into account contributions from kinetic, thermal, and magnetic energy, as well as deviations from axisymmetry.

To define the radial extension of the jet head in the different models, on the north and south sides, respectively, we follow a procedure similar to that described in \cite{Dreas2025}. Specifically:

- first, we calculate $E_\mathrm{sum}=E_\mathrm{kin}+E_\mathrm{th}+E_\mathrm{mag}$ \GREEN{above} spherical surfaces of different radii, for the output times immediately before and after 3\,s after launch;\footnote{To this end, we extend our simulations 50\,ms beyond $t\!=\!3$\,s after launch, calculating $E_\mathrm{sum}$ at $t-50$\,ms and $t+50$\,ms.}

- second, we compute the finite difference of $E_\mathrm{sum}$ over the total time interval, yielding the corresponding flux across each spherical surface at 3\,s after launch;

- third, we define the radial extension of the jet head as the range, from $r_\mathrm{in}$ to $r_\mathrm{out}$, where the above energy flux is larger than 15~per cent of its maximum value (see \citealt{Dreas2025}). 

The values of $r_\mathrm{in}$ and $r_\mathrm{out}$ calculated for the different models are listed in Table~\ref{tab:2}. 
All models have $r_\mathrm{in}$ between about 6 and $7\times10^5$\,km, except for model (b), where $r_\mathrm{in}\!\approx\! 5\times10^5$\,km. This is due to the longer decay time scale adopted in such a \GREEN{model}, which delays the detachment of the jet from the injection site, leading to a more extended jet tail at a given time since the launch.
On the other hand, $r_\mathrm{out}\approx 9\times10^5$\,km in all models, consistently with the fact that, after the breakout, the upper front of the jet propagates essentially at the speed of \GREEN{light}.\footnote{We note that variations in $r_\mathrm{out}$ of a few per~cent are due to the logarithmic grid spacing set up in the radial direction (Section~\ref{sec:setup}). Specifically, the spacing is $\simeq\!4.4$\,km at $\rexc\!=\!380$\,km while $\simeq\!1\!\times\!10^4$\,km at a radius of $\simeq\!9\!\times\!10^5$\,km.}

\begin{figure*}
    \centering    
    \vspace{0.5cm}
    \includegraphics[width=2\columnwidth]{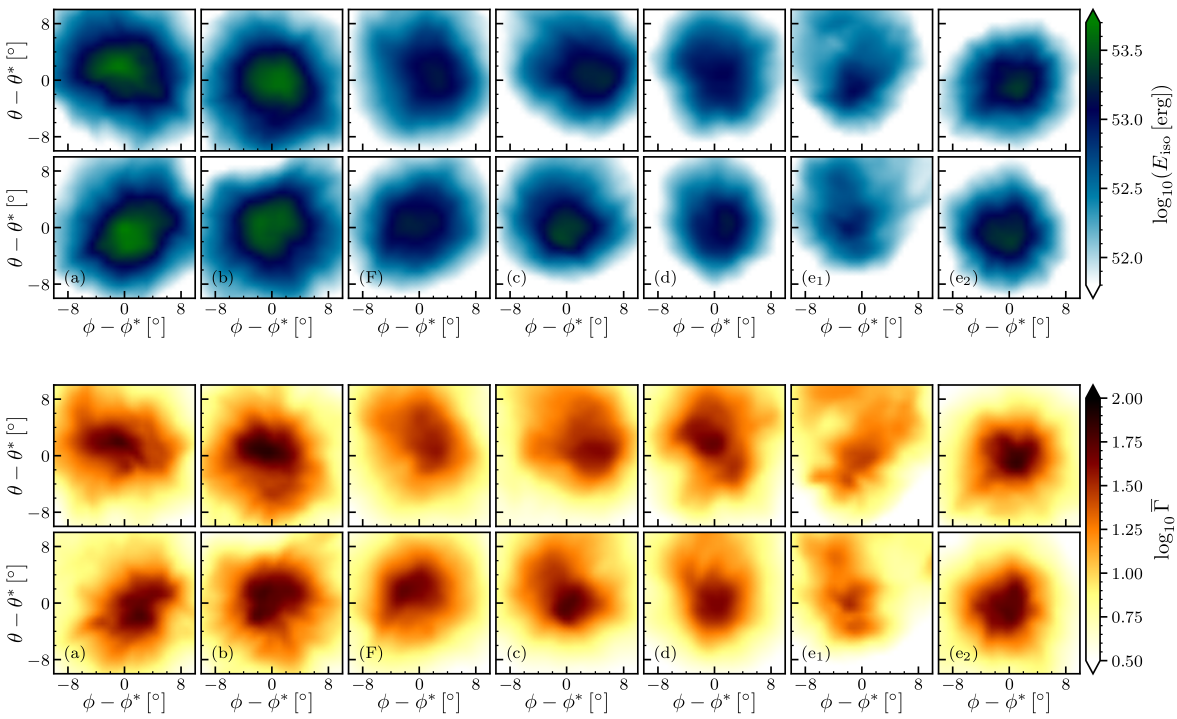}
    \caption{Angular profiles at the jet head of isotropic-equivalent energy (upper set of panels) and radially averaged Lorentz factor (lower set of panels) at 3\,s after launch for the different models. In each set, the top and bottom rows correspond to the north and south sides, respectively. The profiles, given in terms of the spherical angular coordinates ($\theta, \phi$), are centered on the angular location of the injection axis ($\theta^*, \phi^*$).}
    \label{fig:fig9}
\end{figure*}

Once we have determined the radial extension of the jet head at 3\,s after launch, the corresponding angular structure can be described in terms of isotropic-equivalent energy and radially averaged Lorentz factor, defined respectively as \citep[e.g., Eqs.~1-2 in][]{Salafia2022}
\begin{align}\label{eq:Eiso} 
    E_\mathrm{iso}(\theta,\phi) \equiv 4\pi\frac{\mathrm{d}E_\mathrm{sum}}{\mathrm{d}\Omega} = 4\pi\int^{r_\mathrm{out}}_{r_\mathrm{in}}e_\mathrm{sum}(r,\theta,\phi)r^2\mathrm{d}r\,,
\end{align}
and 
\begin{equation}\label{eq:GmMean}
    \overline{\Gamma}(\theta,\phi) = \frac{\int^{r_\mathrm{out}}_{r_\mathrm{in}}\Gamma(r,\theta,\phi)e_\mathrm{sum}(r,\theta,\phi)r^2\mathrm{d}r}{\int^{r_\mathrm{out}}_{r_\mathrm{in}}e_\mathrm{sum}(r,\theta,\phi)r^2\mathrm{d}r}\,,
\end{equation}
where \GREEN{$e_\mathrm{sum}=e_\mathrm{kin}+e_\mathrm{th}+e_\mathrm{mag}$}, and  $(\theta,\phi)$ are the polar and azimuthal angles measured with respect to the spherical polar axis (orthogonal to the jet injection \GREEN{axis}).
The profiles of $E_\mathrm{iso}$ and $ \overline{\Gamma}$ for the different models are shown in Figure~\ref{fig:fig9}, in the upper and lower set of panels, respectively (in each set, the top and bottom rows refer to the north and south sides, respectively).
The maximum values on each side, $E_\mathrm{iso,max}$ and $\overline{\Gamma}_\mathrm{max}$, are listed in Table~\ref{tab:2}.

Compared to our fiducial model (F) (third column panels from the left in Figure~\ref{fig:fig9}),
increasing the initial jet luminosity (model (a); first column) or decay time scale (model (b); second column), thus the energy injected until 3\,s after launch ($E_\mathrm{inj}$), results in a jet head with a core of higher $E_\mathrm{iso}$ and $\overline{\Gamma}$.
Moreover, adopting the same $E_\mathrm{inj}$ in models (a) and (b) \GREEN{results} in similar jet head energetics. In particular, $E_\mathrm{iso,max}$ and $\overline{\Gamma}_\mathrm{max}$ reach up to, respectively, $4.96\times10^{53}\,\mathrm{erg}$ and $62.2$ in model (a), $4.13\times10^{53}\,\mathrm{erg}$ and $75.6$ in model (b), \GREEN{while} $1.64\times10^{53}\,\mathrm{erg}$ and $52.1$ in model (F).

The above comparison can be further explored by considering the ratios $E_\mathrm{sum}/E_\mathrm{inj}$ and $E_\mathrm{kin}/E_\mathrm{sum}$ listed in Table~\ref{tab:2}. 
At 3\,s after launch, in both models (a) and (b) the jet head retains \BUG{about 60~per cent} of $E_\mathrm{inj}$, with \RED{approximately} $90$~per cent of $E_\mathrm{sum}$ being in kinetic form.
This means that \BUG{$\approx40$~per cent} of the injected energy has been spent on jet drilling (and deposited into the surroundings), with most of the remainder being converted into jet acceleration. 
In model (F), \RED{on the other hand,} the jet head features \BUG{$E_\mathrm{sum}/E_\mathrm{inj}\!\approx\!50$~per cent}, while the fraction of kinetic energy is similar to that in models (a) and (b).
We conclude that the lower energy injection in our fiducial model leads to (i) larger energy losses during the earlier propagation stages, which further reduces the final jet power, and (ii) a reduced Lorentz factor, even though the long-term conversion of energy in kinetic form is similar.

Model (c), featuring the same initial jet luminosity but a higher magnetization compared to model (F), shows $E_\mathrm{iso}$ and $\overline{\Gamma}$ distributions that are intermediate between model (F) and models (a) and (b). 
\RED{In model (d),} the higher jet magnetization compensates for the lower \RED{initial} luminosity, yielding a final outcome that is comparable to model (F) (cfr. third and fifth column panels in \GREEN{Figure~\ref{fig:fig9}}).

Focusing now on model (e1) (sixth column panels in Figure~\ref{fig:fig9}), we note that the jet head features a relatively low $E_\mathrm{iso}$ and, notably, the absence of a high-$\overline{\Gamma}$ core (\GREEN{instead} present in all the other models).
\GREEN{This} is consistent with stronger jet-environment interactions in earlier propagation stages \citep[][]{Geng2019}, which, in model (e1), are due to lower initial jet \RED{luminosity and} magnetization within the same realistic environment (\GREEN{as shown in} Section~\ref{sec:choking_tc}).

Reducing the launch time by nearly a factor 2, as in model (e2), leads to a very different configuration after 3\,s. 
Specifically, despite having identical jet parameters as model (e1), model (e2) results in a more relativistic and energetic jet head, with a high-$\overline{\Gamma}$ core (cfr.~last two column panels in Figure~\ref{fig:fig9}).
Notably, $\overline{\Gamma}_\mathrm{max}$ and $E_\mathrm{iso,max}$ \RED{reach up to} 70.6 and $2.27\times10^{53}\,\mathrm{erg}$, respectively, more than twice the corresponding values in model (e1) (Table~\ref{tab:2}).
\RED{This is consistent with reduced jet energy losses in earlier propagation stages.}
In particular, $E_\mathrm{sum}/E_\mathrm{inj}$ in model (e2) approaches \RED{the values observed in models~(a) and (b).}
\begin{table*}
    \centering
    \caption{
    Jet head properties at 3\,s after launch. For each model, labeled as in the first column on the left, we list the inner and outer radial boundaries defining the jet head
    ($r_\mathrm{in}$ and $r_\mathrm{out}$, respectively), along with the maxima of isotropic-equivalent energy $E_\mathrm{iso,max}$ (Eq.~\ref{eq:Eiso}) and radially averaged Lorentz factor $\overline{\Gamma}_\mathrm{max}$ (Eq.\ref{eq:GmMean}). In addition, in the last three columns on the right, we list the ratio between $E_\mathrm{sum}=E_\mathrm{kin}+E_\mathrm{th}+E_\mathrm{mag}$ \GREEN{and} energy injected until 3\,s after launch ($E_\mathrm{inj}$), the ratio between $E_\mathrm{kin}$ and $E_\mathrm{sum}$, and the angular distance from the injection axis containing 50~per~cent of $E_\mathrm{sum}$ ($\Theta_{50}$). See text for further details and discussion.
    }
    \label{tab:2}
    \begin{tabular*}{2.084\columnwidth}{@{\extracolsep{\fill}}lccccccc@{}}
        \hline
        \hline
        \noalign{\vskip 1mm}
        Model & $r_\mathrm{in}\,[10^{5}\,\mathrm{km}]$ & $r_\mathrm{out}\,[10^{5}\,\mathrm{km}]$ & $E_\mathrm{iso,max}\,[10^{53}\,\mathrm{erg}]$ & $\overline{\Gamma}_\mathrm{max}$ & $E_\mathrm{sum}/E_\mathrm{inj}$ & $E_\mathrm{kin}/E_\mathrm{sum}$ &
        $\Theta_\mathrm{50}\,[\degr]$ \\
 	\noalign{\vskip 0.6mm}
 	\hline
 	\noalign{\vskip 1mm} 
 	(a) & 6.37 (6.51) & 8.97 (8.97) & 4.50 (4.96) & 62.2 (60.0) & \BUG{0.60 (0.59)} & 0.90 (0.88) & 5.34 (4.71) \\
        (b) & 5.18 (4.95) & 8.77 (8.77) & 4.13 (3.64) & 75.6 (62.4) & \BUG{0.59 (0.58)} & 0.86 (0.87) & 4.92 (4.87) \\
        (F) & 6.22 (6.37) & 8.77 (8.77) & 1.52 (1.64) & 37.2 (52.1) & \BUG{0.50 (0.49)} & 0.92 (0.92) & 5.81 (5.04) \\
        (c) & 6.37 (6.51) & 8.97 (8.97) & \BUG{1.82 (2.49)} & \BUG{40.8 (58.6)} & \BUG{0.51 (0.50)} & \BUG{0.89 (0.91)} & \BUG{5.80 (4.44)} \\
        (d) & 6.37 (6.37) & 8.77 (8.77) & 1.19 (1.52) & 50.6 (43.2) & \BUG{0.48 (0.42)} & 0.89 (0.92) & 5.28 (4.64) \\
        (e1) & 6.08 (5.94) & 8.77 (8.77) & 1.10 (0.93) & 27.8 (32.6) & \BUG{0.44 (0.46)} & 0.90 (0.93) & 6.58 (7.46) \\
        (e2) & 6.82 (6.82) & 9.18 (9.18) & 2.27 (2.21) & 70.6 (56.3) & \BUG{0.61 (0.61)} & 0.92 (0.91) & 4.41 (4.16)\\
        \noalign{\vskip 0.5mm}  
        \hline
        \hline
        \noalign{\vskip 0.5mm}
    \end{tabular*}
    {\raggedright \textit{Note}: values outside (inside) round brackets refer to the north (south) side. \par}
\end{table*}

\subsection{Angular structure}
\label{sec:structure}

Here, we examine the angular structure at the jet head in detail, focussing on deviations from axisymmetry, particularly in model (a).
We also quantify the jet head's opening angle and compare it across the different models.

\begin{figure}
    \centering    
    \includegraphics[width=0.95\columnwidth]{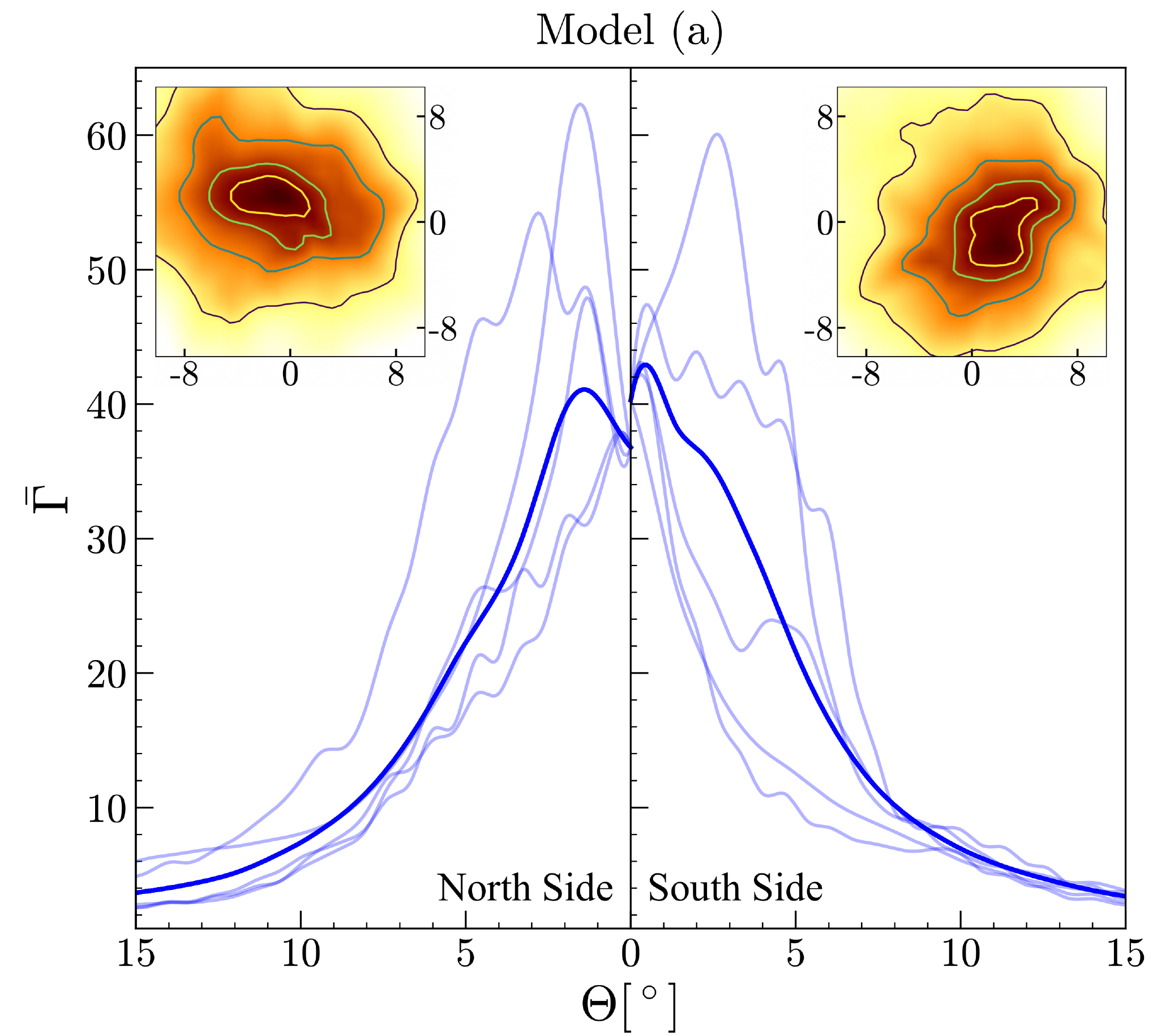}
    \caption{1D profiles of $\overline{\Gamma}$ at varying angular distance from the jet injection axis in model (a), computed at different azimuthal angles from the same axis (thin blue lines) or averaged over the azimuthal direction (thick blue lines), for both the north and south sides.
    In the insets, we show the corresponding 2D profiles of $\overline{\Gamma}$ as in Figure~\ref{fig:fig9}, with isocontours to highlight deviations from axisymmetry. See text for additional details and discussion.}
    \label{fig:fig10}
\end{figure}

Figure~\ref{fig:fig10} shows 1D profiles of $\overline{\Gamma}$ as a function of angular distance from the jet injection axis  ($\Theta$; see Section~\ref{sec:setup}), calculated at different azimuthal angles from the same axis (i.e. $\Phi=60\degr$, $120\degr$, $180\degr$, $240\degr$, and $300\degr$; thin blue lines) or averaged over the azimuthal direction (i.e.~$\overline{\Gamma}$ axisymmetrized with respect to the injection axis; thick blue lines), for both the north and south sides.
Additionally, the insets display the corresponding 2D profiles from Figure~\ref{fig:fig9}, with isocontours highlighting their shape.

We observe large variations in $\overline{\Gamma}$, particularly for $\Theta<10\degr$, when comparing among profiles at different azimuthal angles.
Moreover, the maximum values in the north and south sides ($\overline{\Gamma}_\mathrm{max}=62.2$ and 60.0, respectively), significantly larger than the $\Phi$-averaged maxima ($\approx\!40$ on both sides), occur a few degrees away from the injection axis.
This is the result of the significant deviations from axisymmetry in model (a), as shown in the insets.
As it can be anticipated by looking at Figure~\ref{fig:fig9}, similar conclusions apply to all our models.

Asymmetries found in our 3D simulations prevent reliable fitting of $\overline{\Gamma}$ and $E_\mathrm{iso}$ distributions using axisymmetric analytical profiles, such as those that have been employed to fit afterglow data from GRB\,170817A (e.g., \citealt{Ghirlanda2019}; see also \citealt{Salafia2019}).
In particular, this precludes describing the jet head's angular structure in terms of `standard' parameters, like the half-opening angle $\theta_\mathrm{c}$ of the energetic, high-$\Gamma$ core of the jet (see cited works).

In order to give an effective measure of the jet core opening angle that is applicable in presence of asymmetries, thus allowing for a comparison among our different models, we introduce the parameter $\Theta_{50}$, defined as follows: 
considering the energy $E_\mathrm{sum}$  within the jet head (in the north and south sides, respectively), $\Theta_{50}$ is the angular distance from the injection axis enclosing 50\,per cent of this energy.

The last column on the right in Table~\ref{tab:2} lists $\Theta_{50}$ values calculated for the different models.
Notably, these values vary by only about $1\degr$ (in the range $\approx \BUG{4.4\degr}-5.8\degr$) among the first five models, corresponding to jets launched at the same time (\RED{relative to the} merger) and successfully breaking \GREEN{out}.
Conversely, when launch time is reduced, resulting in less expanded and massive environment, as in model (e2), the jet retains higher collimation ($\Theta_{50}\!\approx\!4\degr$ for this model). 
Finally, in model (e1), in which the jet is significantly choked by the dense surrounding environment, a less collimated angular structure is achieved ($\Theta_{50}\!\approx\!7\degr$).
This reflects the less energetic and less peaked $E_\mathrm{iso}$ distribution (Figure~\ref{fig:fig9}), consistent with a marginally successful \GREEN{breakout} (\citealt{Geng2019,Murguia2021}).\footnote{We remark that these results on $\Theta_{50}$ originate from a fixed half-opening angle at injection, $\Theta_\mathrm{j}=10\degr$ (Table~\ref{tab:1}). Impact of a different $\Theta_\mathrm{j}$ on the final $\Theta_{50}$ will be addressed in future work.} 

For a tentative comparison with the jet's half-opening angle estimated for GRB\,170817A by, e.g., \citet[][]{Ghirlanda2019}, we note that the assumed axisymmetric analytical profile for $E_\mathrm{iso}$, along with the given best fitting parameter set (with $\theta_\mathrm{c}\!\simeq\!3.4\degr$), would result in $\Theta_{50}\!\simeq\!2.9\degr$.
This is a few degrees smaller than in our models. Such a difference
is primarily due to the offset of the peak values with respect to the injection axis in the \GREEN{distributions} shown in Figure~\ref{fig:fig9} (see also Figure~\ref{fig:fig10}).
Indeed, subtracting these offsets from $\Theta_{50}$ yields values around $\approx\!3\degr$. \RED{Furthermore, we remark} that any deviation from axisymmetry, which naturally develops during the jet's evolution, is expected to enlarge $\Theta_{50}$ compared to purely axisymmetric jets.

\section{\RED{Summary and outlook}}
\label{sec:conclusions}

We have presented the first set of 3D RMHD simulations exploring the impact of injection parameters on GRB jet propagation through `realistic' BNS merger environments (i.e.~directly imported from the outcome of a GRMHD BNS merger simulation).
These simulations exploit the numerical framework developed in \citetalias{Pavan2023}, \GREEN{but employing} a refined numerical setup, which ensures higher effective resolution compared to \citetalias{Pavan2023} and correct north-south symmetry in the jet injection prescription (\RED{Section~\ref{sec:setup}; see also} Appendix~\ref{app:improvements}).
\GREEN{Specifically}, we have evaluated the impact of initial jet luminosity and decay time scale, magnetic field contribution ($\Sigma_\mathrm{j}$; see Eq.~\ref{eq:Lmag_tot}), and launch \GREEN{time}.
We have performed simulations in which each of these parameters is varied \GREEN{relative} to a fiducial model, comparing the resulting jet evolution in terms of dynamics, energetics, and final angular \GREEN{structure}.

The results of our investigation can be summarized as follows:

\textbf{Fiducial model.} 
Model (F) describes the evolution of a GRB jet launched at 385\,ms after merger, with \BUG{$5.37\times10^{51}$\,erg/s} initial lumin\-osity, \BUG{$\Sigma_\mathrm{j}\!=\!1.20\times10^{-2}$}, and 300\,ms decay time scale (Subsection~\ref{sec:fiducial}; Section~\ref{sec:angular3s}).

\textit{Dynamics \& Energetics.}
We observe prompt propagation of the jet through the realistic post-merger environment, marked by intense interaction and energy exchange between the jet and its surroundings.
As a result, the jet breaks out at a relatively late time ($\simeq\!275$\,ms after launch), with \BUG{$\approx\!50$} per cent of the injected energy that is deposited into the surroundings, \GREEN{and} a maximum Lorentz factor of $\Gamma_\mathrm{max}\simeq\!60$ \GREEN{reached} at 1\,s after launch.

Comparison with analytical modeling by \cite{Gottlieb2022b} shows very good agreement in describing the jet's advancement into the post-merger environment, when the latter is modeled with a typical rest-mass density power-law index of $\alpha\!=\!4$.
However, significant discrepancies are found in the jet breakout time, with the analytical prediction being up to $\simeq\!150$\,ms longer than our simulation.
We attribute this difference to the lower expansion rate of the realistic BNS merger environment compared to the simple homologous model assumed in the cited work.

\textit{Final angular structure.}
By the end of the simulation (i.e.~3\,s after launch), the fiducial model results in distributions of $E_\mathrm{iso}$ and $\overline{\Gamma}$ at the jet head (see Eqs.~\ref{eq:Eiso}-\ref{eq:GmMean}) with maximum values of $\simeq1.6\times10^{53}$\,erg and 52, respectively. 
Additionally, 50 per cent of the jet's energy is located within an angle of $\Theta_{50}\lesssim6\degr$ from the injection axis. 
A few-degree offset from the injection axis and deviations from axisymmetry are also observed, arising from the 3D evolution of the jet in the realistic post-merger environment.

\textbf{Luminosity and decay time scale.} 
We have examined the impact of increasing the initial jet luminosity (model (a)) and decay time scale (model (b)) by a factor $5/3$ with respect to model (F) (Subsection~\ref{sec:L_tau}; Section~\ref{sec:angular3s}).

\textit{Dynamics \& Energetics.}
The higher initial jet luminosity leads to more efficient drilling through the same realistic environment, resulting in faster propagation and, consequently, a shorter breakout time ($\simeq\!75$\,ms earlier compared to model (F)). 
In contrast, a longer decay time scale has negligible impact on jet propagation or breakout time (the latter being the same as model (F)).
On the other hand, in terms of energetics, jet evolution is similar in models (a) and (b), and enhanced with respect to model (F). 
Specifically, we observe a reduced amount \BUG{($\approx40$ per cent)} of injected energy being deposited into the surroundings, \GREEN{and} $\Gamma_\mathrm{max}\approx100$ at 1\,s after launch. 

\textit{Final angular structure.}
By the end of the simulation, the higher initial jet luminosity or decay time scale result in a much heightened jet head, with $E_\mathrm{iso}$ and $\overline{\Gamma}$ up to a factor of $\simeq\!3$ and $\simeq\!1.5$ higher than in model (F), respectively.
Conversely, we observe less significant impact on the final degree of collimation ($<\!24$ per cent $\Theta_{50}$ variation from model (F)).

\textbf{Magnetic field contribution.} 
We have explored the impact of a higher $\Sigma_\mathrm{j}$ \BUG{(i.e.~equal to $1.56\times10^{-2}$)} while maintaining (model (c)) or reducing \RED{(model (d)) the initial jet luminosity relative to model (F)} (Subsection~\ref{sec:magn_content}; Section~\ref{sec:angular3s}).

\textit{Dynamics \& Energetics.}
The higher initial jet magnetization $\sigma$ (by up to a factor $\simeq\!3$ compared to model (F); Eq.~\ref{eq:sigma_small}) \RED{is associated with better preservation of the initial jet toroidal magnetic field, due to the suppression of hydrodynamical instabilities at the jet-environment interface. 
This is consistent with previous results by \cite{Mignone2010,Gottlieb2020} and is generalized here to the case of realistic BNS merger environments.}
\BUG{Nevertheless, we do not observe significant differences in breakout time, thus in the average advancement speed at the jet front, between models (c) and (F).}
On the other hand, the lower initial jet luminosity of model (d) \RED{makes jet drilling harder, resulting in a breakout time that is $\simeq25-50$\,ms longer compared to model (F).}

\textit{Final angular structure.}
At 3\,s after launch, model (c) shows distributions of $E_\mathrm{iso}$ and $\overline{\Gamma}$ that are intermediate between model (F) and models (a) and (b), indicating that \RED{a higher $\Sigma_\mathrm{j}$} enhances the final jet configuration, but not as much as \RED{a higher injection} energy.
On the other hand, model (d) shows distributions similar to model (F), with the higher $\Sigma_\mathrm{j}$ compensating for the lower luminosity.

\textbf{Chocking condition and launch time.}
The impact of a different launch time has been evaluated comparing the evolution of a \RED{jet with relatively low luminosity and magnetization} \BUG{($L_\mathrm{j}\!=\!3.69\times10^{51}$\,erg/s and $\Sigma_\mathrm{j}=0.46\times10^{-2}$)} launched at 385\,ms (model (e1)) and 185\,ms (model (e2)) relative to the merger (Section~\ref{sec:choking_tc}).
 
In model (e1), \RED{jet drilling is} so inefficient that \BUG{over} half of the injected energy is deposited into the surroundings, and the initial jet collimation and magnetic field configuration are severely compromised shortly after \RED{the} launch.
As a result, the final angular structure lacks a high-$\overline{\Gamma}$ core, with 50 per cent of the energy being located within $\sim\!7\degr$ from the injection axis.

Reducing the launch time by nearly a factor 2, as in model (e2), results in a significantly different outcome.
Specifically, the interaction of the same incipient jet as in model (e1) with a less expanded and less massive surrounding environment leads to a successful jet breakout, with \BUG{less than 40 per cent} of the injected energy \GREEN{being} deposited into the surroundings.
Finally, at 3\,s after launch, a high-$\overline{\Gamma}$ core and higher collimation are observed compared to model (e1), with 50 per cent of the energy being located within $\sim\!4\degr$ from the injection axis.

\vspace{12pt}
Future work will be devoted to extending the above investigation to a larger number of injection parameters, including initial jet opening angle, poloidal-to-toroidal magnetic field ratio, and both initial and terminal Lorentz factors. Additionally, a variety of realistic BNS merger environments \GREEN{will be considered}.
Moreover, we plan to study the effects of a physical resistivity, as already done for simpler analytical BNS merger environments in \cite{Mattia2024}.

By applying the methods developed in \cite{Dreas2025} to our present and future simulations, it will then be possible to further follow the 3D jet evolution on much longer time scales (tens of seconds). This will allow for solid predictions on the final jet configuration shaping the EM \GREEN{emission}.

As final note, we remark that so far we only considered incipient jets that are manually injected into the surrounding post-merger environment. This remains a key limitation that needs to be overcome by considering reference BNS merger simulations where an incipient jet is consistently launched.

\section*{Acknowledgements}

\RED{We thank the anonymous referee for the constructive comments.}
This work was supported by the Italian Ministry of Foreign Affairs and International Cooperation (MAECI), grant number US23GR08.
AP, RC, and ED acknowledge further support by the European Union under NextGenerationEU, via the PRIN 2022 Project “EMERGE: Neutron star mergers and the origin of short gamma-ray bursts", Prot. n.~2022KX2Z3B (CUP~C53D23001150006). 
JVK gratefully acknowledges financial support from CCRG through grants from NASA (Grant No.~80NSSC24K0100) and National Science Foundation (Grants No.~PHY-2110338, No.~PHY-2409706, No.~AST-2009330, No.~OAC-2031744, No.~OAC-2004044). 
Simulations were performed on the Discoverer (Sofia Tech Park, Bulgaria) and GALILEO100 (CINECA, Italy) HPC clusters.
We acknow\-ledge EuroHPC Joint Undertaking for awarding us access to Discoverer via the Regular Access allocations EHPC-REG-2022R03-218 and EHPC-REG-2023R03-160, and CINECA for the availa\-bility of high performance computing resources and support on GALILEO100 through an award under the ISCRA initiative (Grant IsB27\_BALJET). 

\section*{Data Availability}

The data underlying this article will be shared on reasonable request to the corresponding authors.



\bibliographystyle{mnras}




\appendix

\section{Setup improvements}
\label{app:improvements}
\begin{figure*}
    \centering    
    \includegraphics[width=1.5\columnwidth]{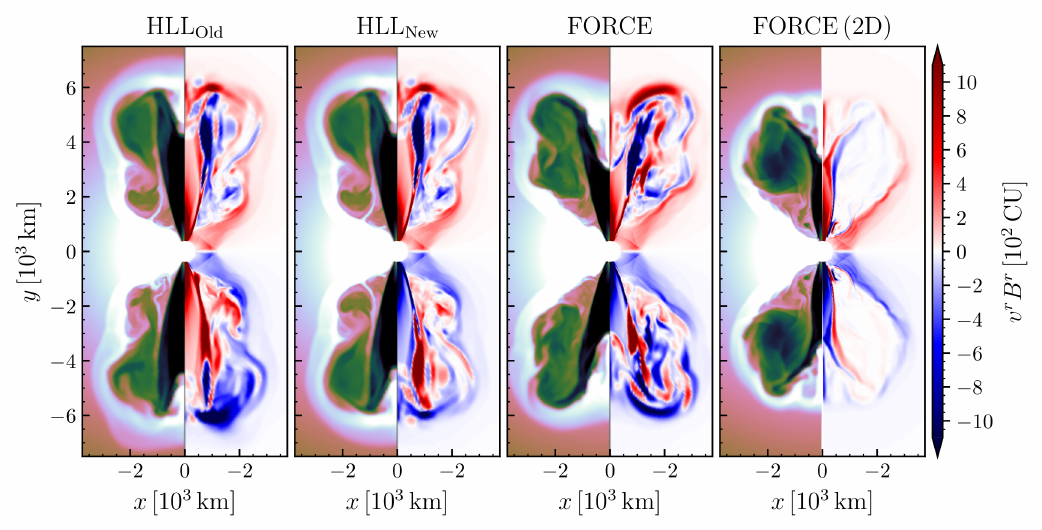}
    \caption{Magnetized GRB jet propagating in a magnetized analytical post-merger environment, adopting a similar numerical setup as \citetalias{Pavan2023} (leftmost panel), including refined implementation of the jet magnetic field \RED{(second panel from left)}, and further replacing the HLL Riemann solver with FORCE \RED{(third panel). In the rightmost panel, moreover, the analogous 2D of the latter case is shown.} In each panel, on the left side, the configuration is shown in terms of the rest-mass density distribution (same colormap as in Figure~\ref{fig:fig2}), while on the right side, in terms of $v^rB^r$ (in $10^2$ code units). See text for further details and discussion.}
    \label{fig:figA1}
\end{figure*}

In this Appendix, we provide quantitative discussion of the improvements in our simulation setup and their impact on our results compared to \citetalias{Pavan2023}. Specifically, these improvements include:

(i) Refined implementation of the jet magnetic field at injection, correctly accounting for `equatorial' symmetry relative to the plane orthogonal to the injection axis (Eqs.~\ref{eq:Br}-\ref{eq:Bphi});

(ii) Less diffusive Riemann solver than the Harten–Lax–van Leer
(HLL), namely the First ORder CEntred (FORCE) solver \citep[][and refs.~therein]{Mattia2022}.\footnote{Note that in the proximity of strong shocks (i.e.~$\Delta x |\nabla P|/P>5.0$, where $\Delta x$ is the local grid spacing), we use \texttt{SHOCK\_FLATTENING} set to \texttt{MULTID} as implemented in \texttt{PLUTO} (see the \href{https://plutocode.ph.unito.it/userguide.pdf}{User’s Guide}), allowing the Riemann solver to switch to HLL, increasing the code robustness.}

To quantify their impact, we present here results from 3D simulations of a magnetized GRB jet propagating in a magnetized analytical post-merger environment, in which each improvement is applied sequentially until reaching the numerical setup adopted in this work.
In these simulations, \RED{the jet is set the same as in model~(c) (Table~\ref{tab:1}), while the environment} is modeled with isotropic rest-mass density and pressure distributions, homologous radial velocity and dipole magnetic field configuration as detailed \GREEN{in \cite{Mattia2024}.}

\RED{To highlight the role of dimensionality, we also present 2D simulations where the same environment and jet models are adopted, but axisymmetrized relative to the jet injection axis.}

In Figure~\ref{fig:figA1}, we show the jet-environment system configuration at $\simeq\!65$\,ms after launch, obtained using a similar numerical setup as \citetalias{Pavan2023} (\RED{$\mathrm{HLL_{Old}}$}), with improvement (i) applied (\RED{$\mathrm{HLL_{New}}$}), and both (i) and (ii) applied (\RED{$\mathrm{FORCE}$}). 
\RED{In the rightmost panel, then, we show the analogous 2D of the latter case.} 
In each panel, on the left side, we show the configuration in terms of the rest-mass density distribution, while on the right side, in terms of $v^rB^r$, \RED{in the meridional $xy$-plane.}

Comparing the \RED{$\mathrm{HLL_{Old}}$} and \RED{$\mathrm{HLL_{New}}$} cases, in the latter we note a higher level of north-south symmetry in the rest-mass density distribution. 
\RED{Furthermore,} the opposite values of $v^rB^r$ between the north and south sides in the \RED{$\mathrm{HLL_{New}}$ case (unlike the $\mathrm{HLL_{Old}}$ one)} indicate correctly oriented radial components: outward radial velocity and magnetic field in the north, and outward radial velocity with inward radial magnetic field in the \GREEN{south}.

\RED{Comparing the $\mathrm{HLL_{New}}$ and $\mathrm{FORCE}$ cases}, then, in the latter we observe a higher effective resolution despite the identical computational grid.
\RED{Notably, this allows Kelvin–Helmholtz instabilities at the jet-environment interface to generate physical north-south asymmetries, which are not observed in the $\mathrm{HLL_{New}}$ case.}
This arises because the FORCE Riemann solver outperforms the HLL solver in resolving fluid \GREEN{discontinuities}, as shown by \cite{Mattia2022}.

\RED{Finally, the comparison between the $\mathrm{FORCE}$ case and its analogous 2D reveals major differences. 
Consistent with previous studies \citep[e.g.,][and refs. therein]{Harrison2018,Gottlieb2020}, the imposed axisymmetry in the latter case confines the jet-environment interaction primarily to the front, leaving the collimation shock at the base nearly unperturbed (resulting in a higher jet collimation). 
Moreover, it causes an unphysical sideways deflection of jet material at the front, resembling a pair of antennas (plug instability; see cited works). 
Finally, in the 3D case we note a higher level of MHD turbulence, with multiple and intense reversals of $v^rB^r$, which are nearly absent in the 2D simulation (see also \citealt{Mattia2024}).
} 

\section{Resolution study}
\label{app:res_stud}

Here, we discuss the results of a simulation performed using a similar setup as fiducial model (F) (Subsection~\ref{sec:fiducial}), but double the resolution along the radial direction, i.e., $1536$ cells along $r\in[\rexc,r_\mathrm{max}]$.
We refer to this high-resolution case as model $\mathrm{(F)_{HR}}$.

\begin{figure}
    \centering    
    \includegraphics[width=0.9\columnwidth]{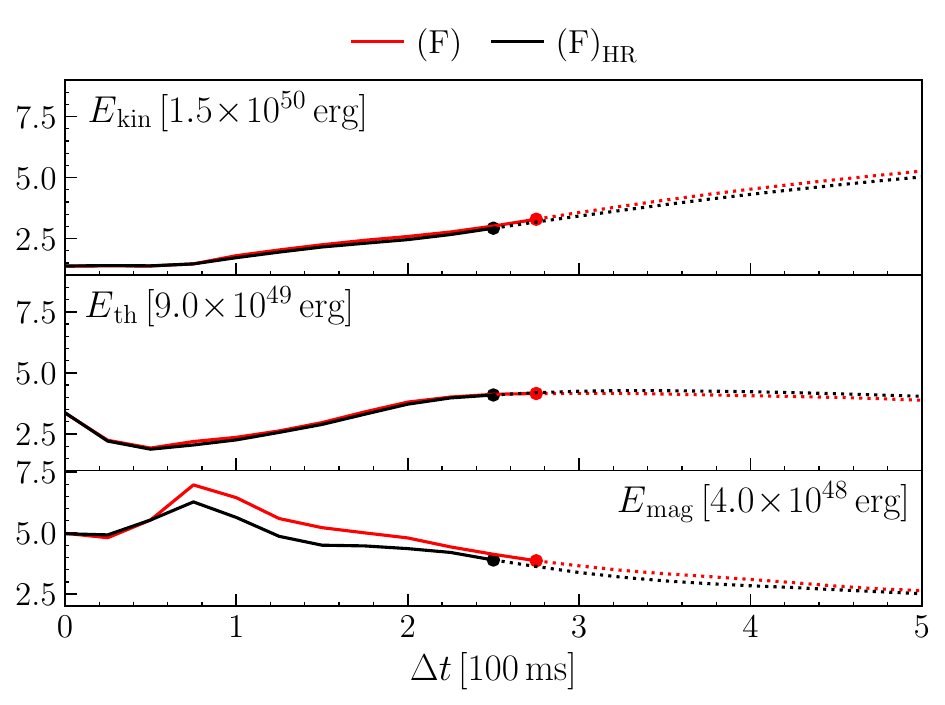}
    \caption{Temporal trends of kinetic ($E_\mathrm{kin}$), thermal ($E_\mathrm{th}$), and magnetic energy ($E_\mathrm{mag}$), from launch until 500\,ms later, for different grid resolution. In particular, model $\mathrm{(F)_{HR}}$ has similar setup as model (F) (Subsection~\ref{sec:fiducial}) while twice the number of grid cells along the radial direction.
    The color dots mark the jet breakout, while post-breakout trends are indicated with dashed lines.
    }
    \label{fig:figB1}
\end{figure}

Figure~\ref{fig:figB1} shows the temporal trends of lab-frame kinetic ($E_\mathrm{kin}$), thermal ($E_\mathrm{th}$), and magnetic energy ($E_\mathrm{mag}$), from launch until 500\,ms later, calculated over the entire computational domain in models (F) and $\mathrm{(F)_{HR}}$ (red and black lines, respectively). For each model, the colored dots mark the jet breakout, while post-breakout trends are indicated with dashed lines. 

We note that $E_\mathrm{kin}$ and $E_\mathrm{th}$ exhibit nearly identical temporal trends across the models, while $E_\mathrm{mag}$ is slightly higher in model (F) starting at $\simeq\!80$\,ms after launch (though this corresponds to only a small variation in the total energy \GREEN{budget}).
Furthermore, the observed differences in breakout time correspond to the 25\,ms interval between consecutive simulation outputs, therefore the actual time difference could be smaller.

We conclude that increasing the resolution has a minor impact on the reproduced evolution in terms of energetics. 
Specifically, \GREEN{the} energy conversion processes during jet propagation in the realistic BNS merger \GREEN{environment result in} nearly identical jet acceleration (\GREEN{or} $E_\mathrm{kin}$ trends). 
Notably, our results remain unaltered up to 500\,ms after launch (beyond the breakout), by which time the structure and energetics of the GRB jet have already been shaped by interactions with the realistic environment.

In addition, the higher $E_\mathrm{mag}$ observed in model (F), at lower 
resolution, indicates that the conversion of magnetic energy into other forms occurs primarily via ideal MHD rather than resistive processes (the latter being of numerical origin in our simulations).
If the latter were dominant, indeed, we would see increasing dissipation of magnetic energy with decreasing resolution (due to the enhanced numerical resistivity), i.e.~lower $E_\mathrm{mag}$ in model (F) compared to $\mathrm{(F)_{HR}}$.
Conversely, when ideal MHD processes are dominant, as in our simulations, magnetic energy is first converted into bulk acceleration; then, interactions between the jet and surrounding environment lead to pressurization, converting part of the kinetic energy gained through MHD acceleration into thermal form.

Finally, we remark that physical resistivity can play a role in GRB jet propagation \citep[e.g.,][]{Mattia2024}. 
Including this within realistic BNS merger environments will be the focus of a follow-up study.


\bsp	
\label{lastpage}
\end{document}